\documentclass[aps,prd,showpacs,floatfix,superscriptaddress,
preprintnumbers,nofootinbib,amssymb,amsmath]{revtex4}
\usepackage{graphicx,bm,color}
\usepackage{subfigure}
\usepackage{slashed}
\newcommand{\id}{{1\!\!1}}
\begin{document}
%%%%%%%%%%%%%%%%%%%%%%%%%%%%%%%%%%%%%%%%%%%%%%%%%%%%%%%%%%%%%%%%%%%%%%%%
%%%%%% TITLE AND AUTHORS %%%%%%%%%%%%%%%%%%%%%%%%%%%%%%%%%%%%%%%%%%%%%%%
%%%%%%%%%%%%%%%%%%%%%%%%%%%%%%%%%%%%%%%%%%%%%%%%%%%%%%%%%%%%%%%%%%%%%%%%
\title{Quark Number Susceptibility : Revisited with
Fluctuation-Dissipation Theorem in mean field theories} 

\author{Sanjay K. Ghosh}
\email{sanjay@jcbose.ac.in}
\affiliation{
Department of Physics \&
Center for Astroparticle Physics and Space Science,
Bose Institute,
Kolkata 700091, India.
}

\author{Anirban Lahiri}
\email{anirban@theory.tifr.res.in}
\affiliation{
Department of Theoretical Physics,
Tata Institute of Fundamental Research,
Mumbai 400005, India.
}

\author{Sarbani Majumder}
\email{sarbani.majumder@saha.ac.in}
\affiliation{
Theory Division, Saha Institute of Nuclear Physics,
Kolkata 700064, India.
}

\author{Munshi G. Mustafa}
\email{munshigolam.mustafa@saha.ac.in}
\affiliation{
Theory Division, Saha Institute of Nuclear Physics,
Kolkata 700064, India.
}

\author{Sibaji Raha}
\email{sibaji@jcbose.ac.in}
\affiliation{
Department of Physics \&
Center for Astroparticle Physics and Space Science,
Bose Institute,
Kolkata 700091, India.
}

\author{Rajarshi Ray}
\email{rajarshi@jcbose.ac.in}
\affiliation{
Department of Physics \&
Center for Astroparticle Physics and Space Science,
Bose Institute,
Kolkata 700091, India.
}
%%%%%%%%%%%%%%%%%%%%%%%%%%%%%%%%%%%%%%%%%%%%%%%%%%%%%%%%%%%%%%%%%%%%%%%%
%%%%%% ABSTRACT, KEYWORDS AND PACS %%%%%%%%%%%%%%%%%%%%%%%%%%%%%%%%%%%%%
%%%%%%%%%%%%%%%%%%%%%%%%%%%%%%%%%%%%%%%%%%%%%%%%%%%%%%%%%%%%%%%%%%%%%%%%
\begin {abstract}
Fluctuations of conserved quantum numbers are associated with the
corresponding susceptibilities because of the symmetry of the system. 
The underlying fact is that these fluctuations as defined through the
static correlators become identical to the direct calculation of these
susceptibilities defined through the thermodynamic derivatives, due to
the fluctuation-dissipation theorem. Through a rigorous exercise we
explicitly show that a diagrammatic calculation of the static
correlators associated with the conserved quark number fluctuations and
the corresponding susceptibilities are possible in case of mean field
theories, if the implicit dependence of the mean fields on the quark
chemical potential are taken into account appropriately. As an aside we
also give an analytical prescription for obtaining the implicit
dependence of the mean fields on the quark chemical potential.

\end{abstract}

\preprint{TIFR/TH/14-21}
\preprint{SINP/TNP/2014/05}
\maketitle
%%%%%%%%%%%%%%%%%%%%%%%%%%%%%%%%%%%%%%%%%%%%%%%%%%%%%%%%%%%%%%%%%%%%%%%%
%%%%%% MAIN BODY %%%%%%%%%%%%%%%%%%%%%%%%%%%%%%%%%%%%%%%%%%%%%%%%%%%%%%%
%%%%%%%%%%%%%%%%%%%%%%%%%%%%%%%%%%%%%%%%%%%%%%%%%%%%%%%%%%%%%%%%%%%%%%%%
%======================================================================%
\section{Introduction}
\label{sc.intro}
%----------------------------------------------------------------------%

In relativistic heavy-ion collisions the fluctuations and correlations
of conserved charges like the baryon number, electric charge
{\it etc.}\ are considered to carry promising signals for the formation
of the exotic Quark Gluon Plasma (QGP). Close to any continuous phase
transition region these are supposed to exhibit critical behavior
\cite{asakawa,jeon_koch,Stephanov_prl,hatta_prd}. Therefore the
characteristics of quark-hadron phase transition can be understood by
analyzing the fluctuations of the system. The fluctuations are often
calculated theoretically through the respective susceptibilities.
For comparison with experimental data various combinations of the
ratios of these susceptibilities constitute important phenomenological
observables \cite{Ejiri,GavGup_11,SG_science}. 

Here we shall discuss the quark number susceptibility (QNS) which
provides the response of the net quark number density to the change
in quark chemical potential. Several first principle studies have been
done to calculate the QNS in various approaches. These include the
numerical simulation of QCD on a space-time lattice (LQCD)
\cite{Gottieb_prl,Gottieb_prd,Alton,Gavai_prd1,Gavai_prd2,
HOTQCD_08,Cheng_09,WB_12,HOTQCD_12} as well as Hard Thermal
Loop (HTL) calculations \cite{Blaizot,purnendu1,purnendu2,jiang,
HTLPT_Najmul,najmul,HMS1_2013,HMS2_2013,HAMSS_2014,HBAMSS_2014}.
In the present manuscript we intend to revisit the calculations of QNS
through the phenomenological models namely the Nambu$-$Jona-Lasinio (NJL)
model and its Polyakov Loop extended version, the PNJL model. We begin
with a brief introduction for the various studies already carried out
within the various QCD inspired models to understand the thermodynamic
properties of strongly interacting matter. 

QCD phase transitions for vanishing and non-vanishing baryon chemical
potential have been studied in a great detail and the possible phases
that may arise in the phase diagram have been addressed
\cite{Fukushima_PLB,BCPR_05,ratti1, rajarshidaA,rajarshidaB,
ratti_diquark,pnjl_pd_SFR,rajarshidaC,Fukushima_PRD,Kahara_2peak,
pnjl_FZL,njl_pd_CRS,CEP_pnjl_KKMY,pnjl_pd_review,pnjl_para_CHRS,
pnjl_QHPT,Sarbani,njl_pd_DR,pnjl_para_IJMP}. Similar studies were
also carried out for imaginary chemical potential
\cite{pnjl_imgnmu_SKKY,pnjl_imgnmu_SKY,pnjl_imgnmu_MSFR} and the
well-known Roberge-Weiss periodicity is discussed in that context.
Although hadrons are not present as dynamical degrees of freedom
in NJL or PNJL model, mesonic modes at real as well as imaginary
chemical potential are studied as collective excitations within Random
Phase Approximation \cite{DA_95,buballa_meson,ratti_meson,paramita_meson,
pnjl_meson_reim2,pnjl_meson_Costa,njl_meson_scrn,pnjl_meson_reim1}
and formation of baryons composed of quarks and diquarks has also been
studied by solving the Dyson-Schwinger equation \cite{pnjl_particle,
pnjl_baryon}. Mesons involving heavy quarks have been studied recently
in NJL and PNJL models \cite{Dmeson_pnjl}. The average phase factor of
QCD determinant is evaluated through PNJL model in
Ref.\cite{avg_phase_pnjl}, where it is argued that since CEP lies
within the region of vanishing phase factor, location of CEP cannot
be determined by LQCD alone. NJL model is explored in the context of
CP restoring phase transition \cite{CP_pnjl}, where it is shown that
nontrivial vacuum term of NJL model can always alter the qualitative
aspects of the high temperature phase transition. The issue of color
neutrality is crucially investigated in Ref.\cite{color_pnjl1,
color_pnjl2}. Interplay between chiral and deconfinement transition
is also investigated with U(1) valued boundary condition for
fermionic fields \cite{dqc_pnjl,dqc_njl_TKM,dqc_njl_Xu}, where
quantities namely dual quark condensate or dressed Polyakov loop
seem to be very effective in those studies. Effects of theta vacuum
on QCD phase structure is investigated using PNJL model in
Ref.\cite{thetavac_pnjl}. NJL and PNJL model are also studied within
background magnetic field \cite{magfld_pnjl_BB,magfld_pnjl_GR,
magfld_pnjl_K,magfld_pnjl_AS}. Another interesting phenomenon namely
chiral magnetic effect, which could possibly explain the observed
charge separation observed in the STAR experiment at RHIC, Brookhaven,
is also investigated through PNJL model \cite{CME1_pnjl,CME2_pnjl}.
Existence of conjectured chirally symmetric but confined phase in QCD
phase diagram which is popularly named as quarkyonic phase is
discussed in terms of PNJL model\cite{AAGNR_2008,qrkync_pnjl_MLRN,
qrkync_pnjl_SKSY,qrkync_pnjl_2013}. Role of axial anomaly and
vector interaction determining the phase diagram of QCD
is studied in Ref.\cite{vecint_pnjl_CNB,axanml_pnjl_RSCS,
axanml_pnjl_PB,vecint_pnjl_estm,vecint_pnjl_BHW}. For three degenerate
or non-degenerate flavored system, it can be shown that NJL vacuum is
unstable unless one incorporates eight-quark interaction. Modification
has been done in this direction for both NJL and PNJL model
\cite{njl_8q_OHP,njl_8q_OHBP,njl_8q_OHMBP,njl_8qpd,paramita_8qPD}.
In general NJL model in its local version is applicable within
a restricted momentum range which is governed by the cutoff parameter,
that appears while regularizing the divergent momentum integrals.
To overcome this limitation a non-local version of the model was
proposed and developed recently \cite{nonlocal_pnjl_SFR,
nonlocal_pnjl_HRCW,nonlocal_pnjl_HKW,nonlocal_pnjl_KHW,
nonlocal_pnjl_PDS}. Furthermore finite volume effect which is
relevant for studying a system created in heavy-ion collisions,
has been studied very recently in Ref.\cite{finitevol_pnjl}.
Various interesting features of Polyakov loop \cite{MAS3_2006,
MAS_12,MAS_14,aminul1} have encouraged people to study within
different formalisms. Inclusion of gluon Polyakov loop is studied in
various aspects \cite{MAS1_2006,MAS2_2006,Muller}. 
Interestingly NJL model has also been studied
within Monte-Carlo framework also \cite{MCNJL1,MCNJL2}. Studies of
various transport coefficients in NJL model framework have also
been reported recently\cite{njl_Marty,etaNJL_Sabyasachi}.

The quark number susceptibility has been studied extensively within
the framework of NJL and PNJL models \cite{rajarshidaA, rajarshidaC,
FLW_2010,sasaki_redlich, ratti_plb, anirban1,fluc_WF,anirban2}.
These studies revealed the order parameter like behavior
of QNS, similar to that obtained in LQCD at vanishing baryon
chemical potential. Recently it has been shown by some of us
\cite{isobreak_pnjl} that when isospin symmetry is broken explicitly,
the baryon-isospin correlations exhibit an almost linear scaling with
the scale of isospin breaking over the entire temperature-baryon chemical
potential phase plane.

Another well-known formulation to study the strongly interacting matter
in nonperturbative regime is the Quark Meson (QM) model and its Polyakov
loop extended version (PQM), which is used to explore phase transition
and phase diagram of QCD \cite{pqm_pd_SPW1,pqm_pd_SPW2,
crit_pqm_SFR,thermodynamics_pqm_SSFR,pqm_pd_HPS, sandeep_vacuum}
as well as quark number susceptibility \cite{Schaefer_2007,kurtosis_pqm,
Schaefer_2010,skokov,qns_mu_pqm,Sandeep,charge_fluc_pqm}.

The QNS is the response of the conserved number density, $\rho$,
with infinitesimal variation in the  chemical potential $\mu +\delta\mu$
as an external source. It is then defined as the second order derivative
of pressure, $\cal{P}$, with respect to $\mu$. On the other hand,
according to the fluctuation-dissipation theorem (FDT), the QNS may
also be obtained from the time-time component of the current-current
correlator in the vector channel \cite{purnendu2,
calen,Kunihiro_PLB,Hatsuda_NJL,Fujii04}. 
The QNS is then expressed as
%%%%%%%%%%%%%%%%%%%%%%%%%%%%%%%%%%%%%%%%%%%%%%%%%%%%%%%%%%%%%%%%%%%%%%%%
\begin{eqnarray}
\chi_q &=&  \frac{\partial\rho}{\partial\mu}
=\frac{\partial^2{\mathcal{P}}}{\partial\mu^2}
=\int d^4x \langle J_0(0,\vec{x})J_0(0,\vec{0})\rangle 
= -\lim_{l\rightarrow 0} {\rm{Re}} \Pi_{00}(0,l)
= \lim_{l\rightarrow 0} \beta
\int_{-\infty}^{+\infty} \frac{d \omega}{2 \pi}
\frac{-2}{1-e^{-\beta\omega}} {\rm{Im}} \Pi_{00}(\omega,l),
\label{eq.fluc_diss_def}
\end{eqnarray}
%%%%%%%%%%%%%%%%%%%%%%%%%%%%%%%%%%%%%%%%%%%%%%%%%%%%%%%%%%%%%%%%%%%%%%%%
where $J_{0}$ is the temporal component of the vector current
and $\Pi_{00}(\omega,l)$ is the time-time component of the vector
correlation function or self-energy with external four momenta
$L\equiv (\omega,l=|\vec l|)$. Because of the symmetry of the system 
the FDT guarantees that the
thermodynamic derivative with respect to the external source, $\mu$
is related to the time-time component of static correlation function
in the vector channel. This is  known as {\it thermodynamic sum rule}
\cite{purnendu2,calen,Kunihiro_PLB}.

In usual perturbation theory the loop expansion and coupling expansion are
symmetric and the thermodynamic consistency is automatic, for a given 
order of coupling $\alpha_s$. So, it really does not matter which of
the equivalent definition is used in (\ref{eq.fluc_diss_def})
to compute QNS for a given order of $\alpha_s$. For resummed approach
like Hard Thermal Loop perturbation theory, the loop expansion and
coupling expansion are not symmetric because higher loops contribute
to the lower order in $\alpha_s$. Unlike usual perturbation theory,
in resummed case an appropriate measure is to be employed
\cite{purnendu2,najmul,HAMSS_2014,HBAMSS_2014} if one desires to
compute QNS in a given order in $\alpha_s$  correctly using
(\ref{eq.fluc_diss_def}). In effective approaches like NJL
or PNJL models the QNS is usually obtained \cite{rajarshidaA, rajarshidaC,
FLW_2010,sasaki_redlich, ratti_plb, anirban1,fluc_WF,anirban2} 
as the second order Taylor coefficient of pressure when it is Taylor
expanded in the direction of the quark chemical potential, $\mu$
with an approximation $\mu< T$. In model calculations any response of a
thermodynamic quantity to some external parameters should also
account for the fact that the mean fields also depend implicitly
on those external parameters \cite{rajarshidaA}. Therefore, a proper
care has to be taken to relate the thermodynamic derivatives
(viz., QNS) with the fluctuation associated with the conserved density.
One of the  purposes of the present work is to demonstrate whether
the effective models like NJL and PNJL explicitly obey the FDT
vis-a-vis thermodynamic sum rule.

Furthermore, the implicit dependence of the mean fields are usually
obtained numerically but this may contain numerical errors which will
increase with the order of derivatives. This is true for either a
direct numerical derivative as well as the method of using a
fitting function for a polynomial expansion of pressure in terms
of chemical potential to corroborate with the Taylor expanded pressure. 
Recently a numerical technique based on
algorithmic differentiation \cite{AD_WWS,AD_CEP,AD_SW} has been
developed to solve this shortcoming. In the present work we also propose
an alternative analytical formalism to calculate the derivatives of
the mean fields with respect to the external parameters. Using these
derivatives we are going to calculate the QNS in a consistent manner.

The paper is organized as follows. In the Sec.~\ref{sc.fldsder} we
discuss the formalism for obtaining the derivatives of the mean
fields in our alternative approach in the NJL and the PNJL models.
In Sec.~\ref{sc.qns_fdt} the calculations of QNS  exploiting the FDT
for a toy model and in effective approaches like NJL as well as in 
PNJL models are presented.  Finally we draw our conclusions in
Sec.~\ref{sc.concl}.

%======================================================================%
\section{Calculating the derivatives of the mean fields}
\label{sc.fldsder}
%----------------------------------------------------------------------%
\subsection{NJL model}
\label{sc.fldsder_njl}
%----------------------------------------------------------------------%
The Lagrangian for the 2 flavor NJL model at finite quark chemical
potential ($\mu_q$) is given as
\cite{Hatsuda_NJL,Klevansky:rmp,buballa},
%%%%%%%%%%%%%%%%%%%%%%%%%%%%%%%%%%%%%%%%%%%%%%%%%%%%%%%%%%%%%%%%%%%%%%%%
\begin{equation}
\mathcal{L}_{\rm NJL}=\bar{\psi}(i\slashed \partial
-m_0 +\gamma_0\mu_q)\psi + \frac{G}{2}
[(\bar{\psi}\psi)^2+(\bar{\psi}i\gamma_5\vec{\tau}\psi)^2].
\end{equation}
%%%%%%%%%%%%%%%%%%%%%%%%%%%%%%%%%%%%%%%%%%%%%%%%%%%%%%%%%%%%%%%%%%%%%%%%
In the mean field approximation the pion condensate 
$\langle\bar{\psi}i\gamma_5\vec{\tau}\psi\rangle=0$ and the 
Lagrangian can be rewritten in terms of the chiral condensate 
$\sigma=\langle\bar{\psi}\psi\rangle=
\langle\bar{u}u+\bar{d}d\rangle=\sigma_u+\sigma_d$ as,
%%%%%%%%%%%%%%%%%%%%%%%%%%%%%%%%%%%%%%%%%%%%%%%%%%%%%%%%%%%%%%%%%%%%%%%%
\begin{equation}
\mathcal{L}_{\rm MF}=\bar{\psi}(i\slashed \partial
-m_0+\gamma_0\mu_q+G\sigma)\psi - \frac{G}{2}\sigma^2.
\label{eq.lag_mf}
\end{equation}
%%%%%%%%%%%%%%%%%%%%%%%%%%%%%%%%%%%%%%%%%%%%%%%%%%%%%%%%%%%%%%%%%%%%%%%%
In the mean field approach, the thermodynamic potential is a functional
of the mean field $\sigma(m_0,T,\mu_q)$, and is given as,
%%%%%%%%%%%%%%%%%%%%%%%%%%%%%%%%%%%%%%%%%%%%%%%%%%%%%%%%%%%%%%%%%%%%%%%%
\begin{equation}
\Omega[\sigma,m_0,T,\mu_q]
=-i\mathrm {Tr}[\ln {S_1}^{-1}]+\frac{G}{2}\sigma^2.
\label{eq.omega}
\end{equation}
%%%%%%%%%%%%%%%%%%%%%%%%%%%%%%%%%%%%%%%%%%%%%%%%%%%%%%%%%%%%%%%%%%%%%%%%
Here and unless stated, \textquoteleft Tr\textquoteright\ denotes the
sum over color, flavor and Dirac indices as well as the four-momentum
and any other notation involving trace operation will be clarified
accordingly. The first term on the right hand side of (\ref{eq.omega})
is the fermionic contribution related to the dressed propagator $S_1$
where,
%%%%%%%%%%%%%%%%%%%%%%%%%%%%%%%%%%%%%%%%%%%%%%%%%%%%%%%%%%%%%%%%%%%%%%%%
\begin{equation}
{S_1}^{-1}=\slashed p -m_0 +\gamma_0\mu_q+G\sigma={S_0}^{-1}+G\sigma,
\label{eq.mod_prop_inv}
\end{equation}
%%%%%%%%%%%%%%%%%%%%%%%%%%%%%%%%%%%%%%%%%%%%%%%%%%%%%%%%%%%%%%%%%%%%%%%%
and $S_0$ is the bare propagator with current quark mass $m_0$.
The second term in $\Omega$ may be considered as the background
contribution of the mean field $\sigma$. To utilize thermodynamic
relations, derivatives of $\Omega$ with respect to the various
parameters and the mean fields are often necessary. In this
regard care has to be taken for the explicit appearances of
the parameters as well as their presence through the mean fields.
For example, the computation of quark number susceptibility
requires derivatives of $\Omega$ with respect to $\mu_q$ that
appear explicitly in $\Omega$ as well as their implicit effects
through $\sigma$ \cite{rajarshidaA}. This is an important observation
that we want to revisit in the present manuscript. Therefore we set the
notation for the explicit derivatives by ${\partial}/{\partial x}$ and
that for the total derivatives with ${d}/{dx}$ for some parameter $x$.

Let us start with the computation of the mean field $\sigma$. One way
to obtain $\sigma$ is to use the stationarity condition
$\partial \Omega/ \partial \sigma =0$ in the mean field approximation,
which gives,
%%%%%%%%%%%%%%%%%%%%%%%%%%%%%%%%%%%%%%%%%%%%%%%%%%%%%%%%%%%%%%%%%%%%%%%%
\begin{equation}
\sigma=i\mathrm {Tr}(S_1).
\label{eq.sig_def}
\end{equation}
%%%%%%%%%%%%%%%%%%%%%%%%%%%%%%%%%%%%%%%%%%%%%%%%%%%%%%%%%%%%%%%%%%%%%%%%
On the other hand, we may also use the defining equation
$\sigma=\partial\Omega/\partial m_0$, which gives the same result as in
Eq.(\ref{eq.sig_def}). However as discussed above, for the derivative with
respect to $m_0$ one should also consider the implicit dependence on $m_0$
of $\Omega$ through $\sigma$. In that case we should rather use the
relation,
%%%%%%%%%%%%%%%%%%%%%%%%%%%%%%%%%%%%%%%%%%%%%%%%%%%%%%%%%%%%%%%%%%%%%%%%
\begin{equation}
\sigma=\frac{d\Omega}{dm_0}=\frac{\partial\Omega}{\partial m_0}+
\frac{\partial\Omega}{\partial\sigma}\cdot\frac{d\sigma}{dm_0}.
\label{eq.sigma_fundef}
\end{equation}
%%%%%%%%%%%%%%%%%%%%%%%%%%%%%%%%%%%%%%%%%%%%%%%%%%%%%%%%%%%%%%%%%%%%%%%%
Using the stationarity condition for the mean field, the second term
vanishes and therefore in this case we get back to the original defining
equation of $\sigma$. Interestingly, if we straightway calculate
$\frac{d\Omega}{dm_0}$ from Eq.(\ref{eq.omega}) and demand it to be equal
to $\sigma$, without imposing the stationarity condition (like in
Eq.(\ref{eq.sigma_fundef})), then Eq.(\ref{eq.sig_def}) will emerge as a
consistency condition.

The transcendental nature of the solutions of (\ref{eq.sig_def})
is apparent. Thus a closed form analytical expression of $\sigma$ as
a function of $m_0$, $T$ and $\mu_q$ cannot be obtained from this equation.
One has to solve the equations numerically and obtain $\sigma(T,\mu_q)$.

The implicit derivatives, in general do not disappear, as can be seen
from the chiral susceptibility, which is the second order derivative
of thermodynamic potential with respect to $m_0$,
%%%%%%%%%%%%%%%%%%%%%%%%%%%%%%%%%%%%%%%%%%%%%%%%%%%%%%%%%%%%%%%%%%%%%%%%
\begin{eqnarray}
\chi_{\sigma}  = \frac{d^2 \Omega}{d m_0^2}
= \frac{\partial^2 \Omega}{\partial m_0^2} +
2\frac{\partial^2 \Omega}{\partial\sigma\partial m_0}
\cdot\frac{d\sigma}{d m_0}+
\frac{\partial^2 \Omega}{\partial\sigma^2}
\cdot\Big(\frac{d\sigma}{d m_0}\Big)^2+
\frac{\partial\Omega}{\partial\sigma}\cdot
\frac{d^2\sigma}{d m_0^2}.
\label{eq.chiral_suscp}
\end{eqnarray}
%%%%%%%%%%%%%%%%%%%%%%%%%%%%%%%%%%%%%%%%%%%%%%%%%%%%%%%%%%%%%%%%%%%%%%%%
The last term in Eq.(\ref{eq.chiral_suscp}) again vanishes due to
stationarity condition of the mean field, but the second and third
terms will remain and give the implicit contributions.
This implicit dependence is what makes life a little difficult in the
mean field calculations which are otherwise quite straightforward.
While the explicit derivatives can be systematically obtained up to
any desired order in a closed analytic form, the implicit contributions
are usually obtained through numerical derivatives. Normally one has to
resort to numerical evaluation of the total derivatives like
$\chi_{\sigma}$, or the implicit part like $d\sigma/d m_0$. Such
derivatives may either be done by direct difference approximations or
by a Taylor series method as proposed by us in Ref.\cite{rajarshidaA}.
Unfortunately both these methods tend to give either large errors or
become quite insensitive as the order of the derivatives are increased.

A possible alternative to these numerical techniques has been explored
in Refs.\cite{AD_WWS,AD_CEP,AD_SW} through the method of algorithmic
differentiation. Derivatives up to very high orders may be computed in
this technique. Though very efficient and less error prone even
for obtaining very high derivatives, the method is algorithmically
involved.

One of our main focuses here is to obtain the implicit contribution in
a semi-analytic approach so that numerical uncertainties are minimized.
Here we shall outline a simple algorithm for obtaining derivatives
in a straightforward semi-analytical procedure. The procedure is
completely analytic as far as obtaining the derivatives go. We shall
argue that the total derivatives at any order can be completely
expressed only in terms of explicit derivatives up to one lower order.
Therefore one can obtain the expressions for total derivatives
completely analytically. Only the values of the final expressions
so obtained are computed numerically.

For this purpose we shall discuss the derivatives with respect to
$\mu_q$, though the methodology would be identical for any other similar
derivatives. The only numerics involved will be the momentum integrals,
and to this end all the methods of differentiation would have identical
efficiency and accuracy.

The derivatives of $\Omega$ with respect to $\mu_q$ would give the quark
number and its susceptibilities. Since the quark number is exactly
conserved one may use the Ward-Takahashi identity to derive the
corresponding three-point functions for the bare and effective theories.
The identity is given as,
%%%%%%%%%%%%%%%%%%%%%%%%%%%%%%%%%%%%%%%%%%%%%%%%%%%%%%%%%%%%%%%%%%%%%%%%
\begin{equation}
q^{\mu}\Gamma_{\mu}(p,p+q)={S_1}^{-1}(p+q)-{S_1}^{-1}(p).
\label{eq.WTI_exact}
\end{equation}
%%%%%%%%%%%%%%%%%%%%%%%%%%%%%%%%%%%%%%%%%%%%%%%%%%%%%%%%%%%%%%%%%%%%%%%%
Here $p$ and $q$ denote the four-momentums of the fermion and the
boson respectively. Now, $q_{\mu}\to 0$ limit of
Eq.(\ref{eq.WTI_exact}) yields the Ward Identity,
%%%%%%%%%%%%%%%%%%%%%%%%%%%%%%%%%%%%%%%%%%%%%%%%%%%%%%%%%%%%%%%%%%%%%%%%
\begin{equation}
\frac{d{S_1}^{-1}}{d p^{\mu}}=\Gamma_{\mu}(p,p),
\label{eq.WI_diff}
\end{equation}
%%%%%%%%%%%%%%%%%%%%%%%%%%%%%%%%%%%%%%%%%%%%%%%%%%%%%%%%%%%%%%%%%%%%%%%%
which in differential form, gives the insertion factor corresponding
to the zero momentum boson line into an internal fermion line.
In the imaginary time formalism, at finite temperature and 
chemical potential, the fourth component of momentum becomes
$p_0=i(2n+1)\pi T+\mu_q$ and thus Eq.(\ref{eq.WI_diff})
can be written as \cite{kapusta},
%%%%%%%%%%%%%%%%%%%%%%%%%%%%%%%%%%%%%%%%%%%%%%%%%%%%%%%%%%%%%%%%%%%%%%%%
\begin{equation}
\frac{d{S_1}^{-1}}{d \mu_q}=\Gamma_{0}(p,p).
\label{eq.WI_diff_finiteT}
\end{equation}
%%%%%%%%%%%%%%%%%%%%%%%%%%%%%%%%%%%%%%%%%%%%%%%%%%%%%%%%%%%%%%%%%%%%%%%%
Using Eq.(\ref{eq.mod_prop_inv}) in the above relation we obtain,
%%%%%%%%%%%%%%%%%%%%%%%%%%%%%%%%%%%%%%%%%%%%%%%%%%%%%%%%%%%%%%%%%%%%%%%%
\begin{equation}
\frac{d{S_1}^{-1}}{d \mu_q}=\gamma_0+\Big(G \frac{d \sigma}
{d \mu_q}\Big) \cdot \id_D \equiv \Gamma_0,
\label{eq.vertex_eff}
\end{equation}
%%%%%%%%%%%%%%%%%%%%%%%%%%%%%%%%%%%%%%%%%%%%%%%%%%%%%%%%%%%%%%%%%%%%%%%%
where $\id_D$ is the identity matrix in Dirac space. For the bare
propagator we get the expected insertion factor for non-interacting
quarks as,
%%%%%%%%%%%%%%%%%%%%%%%%%%%%%%%%%%%%%%%%%%%%%%%%%%%%%%%%%%%%%%%%%%%%%%%%
\begin{equation}
\frac{d{S_0}^{-1}}{d \mu_q}=\gamma_{0},
\label{eq.vertex_bare}
\end{equation}
%%%%%%%%%%%%%%%%%%%%%%%%%%%%%%%%%%%%%%%%%%%%%%%%%%%%%%%%%%%%%%%%%%%%%%%%

\begin{figure} [!ht]
\subfigure[]
{\includegraphics [scale=0.6] {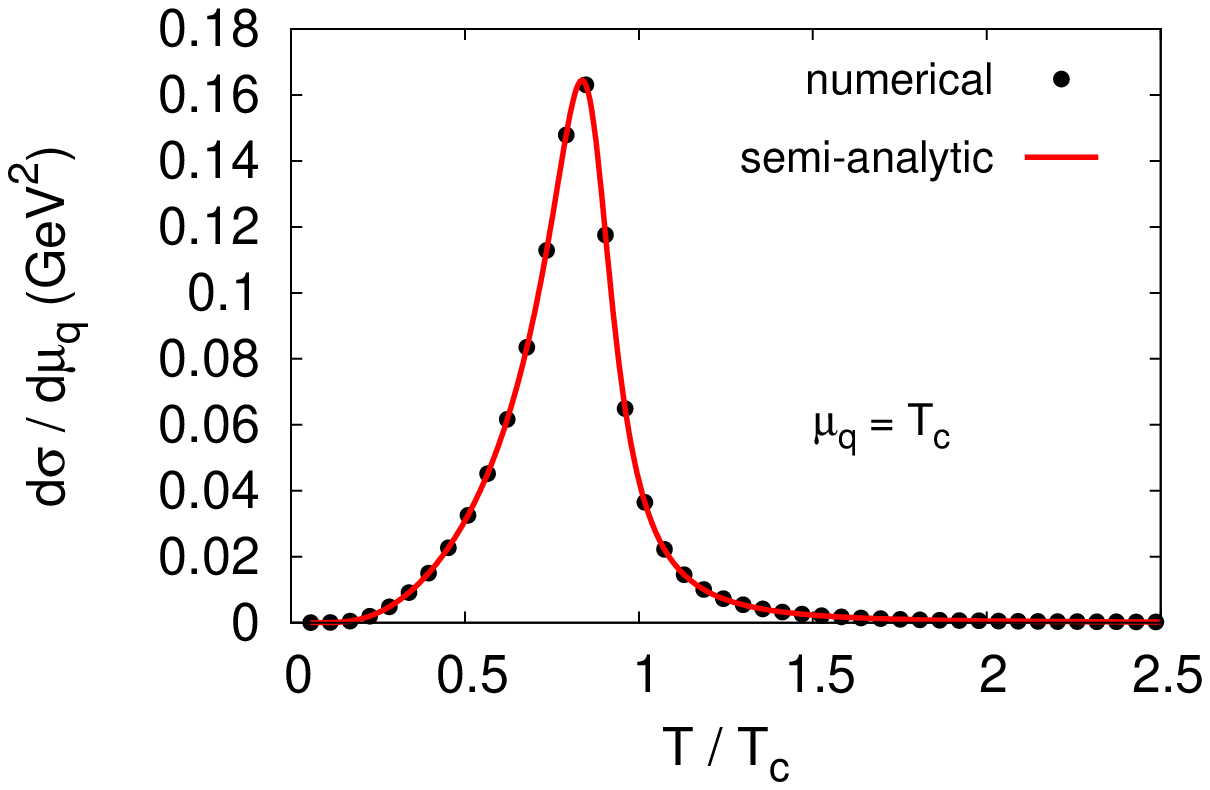}
\label{fg.fldsder1_njl}}
\subfigure[]
{\includegraphics [scale=0.6] {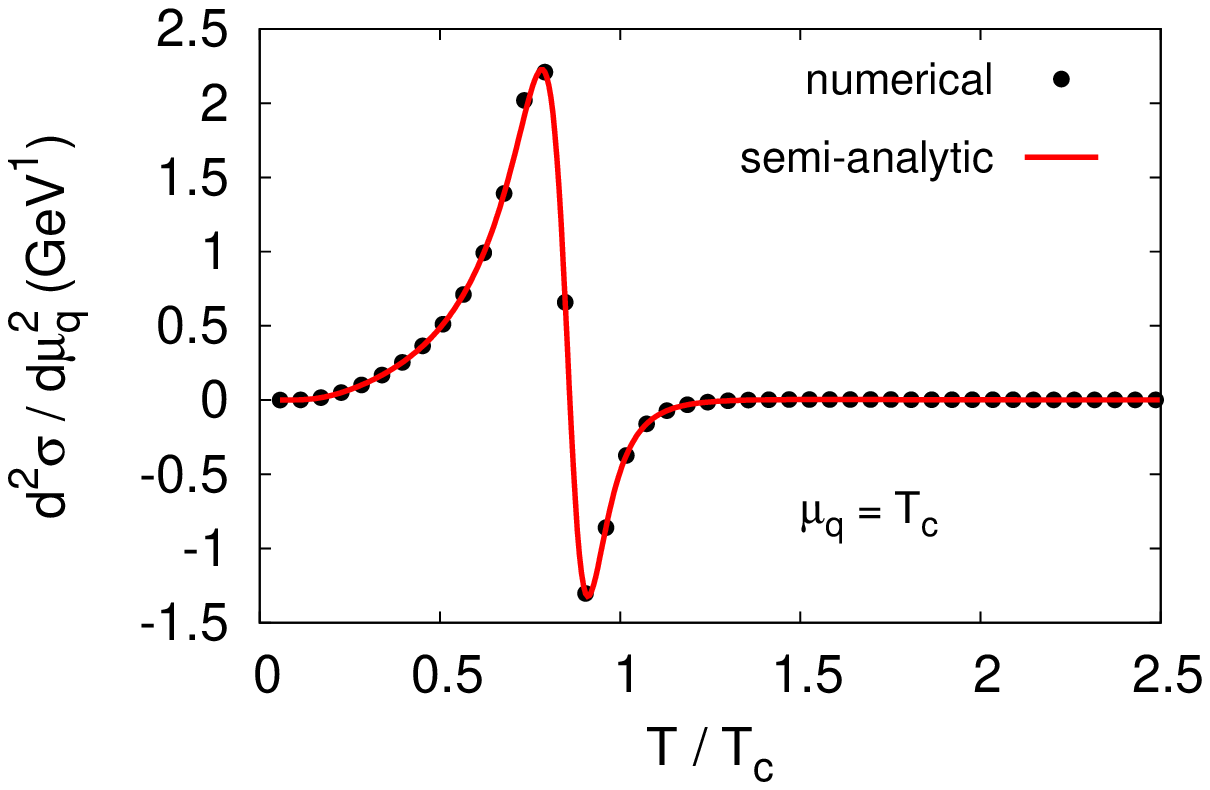}
\label{fg.fldsder2_njl}}
\caption{First (a) and second (b) derivatives of the mean field
$\sigma$ with respect to $\mu_q$ in NJL model at $\mu_q=T_c$. Here
the points represent the result from numerical differentiation and
the lines are from the semi-analytical approach described in the text.}
\label{fg.fldsder_njl}
\end{figure}

Let us now consider the derivative of $\sigma$ from
Eq.(\ref{eq.sig_def}) w.r.t.\ $\mu_q$ which gives,
%%%%%%%%%%%%%%%%%%%%%%%%%%%%%%%%%%%%%%%%%%%%%%%%%%%%%%%%%%%%%%%%%%%%%%%%
\begin{equation}
 \frac{d \sigma}{d \mu_q}=-i\mathrm {Tr}[S_1 \Gamma_0 S_1]
 =-i\mathrm{Tr}(S_1 \gamma_0 S_1)-
 i\mathrm{Tr}\Big(S_1 G\frac{d \sigma}{d \mu_q}S_1\Big),
\label{eq.dsigmadmu_def}
\end{equation}
%%%%%%%%%%%%%%%%%%%%%%%%%%%%%%%%%%%%%%%%%%%%%%%%%%%%%%%%%%%%%%%%%%%%%%%%
where the effective three-point function from Eq.(\ref{eq.vertex_eff})
is used. For the bare propagator $S_0$ one can easily check that
%%%%%%%%%%%%%%%%%%%%%%%%%%%%%%%%%%%%%%%%%%%%%%%%%%%%%%%%%%%%%%%%%%%%%%%%
\begin{equation}
\frac{d S_0}{d \mu_q}=-S_0 \gamma_0 S_0,
\label{eq.unit}
\end{equation}
%%%%%%%%%%%%%%%%%%%%%%%%%%%%%%%%%%%%%%%%%%%%%%%%%%%%%%%%%%%%%%%%%%%%%%%%
which is basically another form of Ward identity for bare three-point
function. The corresponding relation for the dressed propagator $S_1$
with the effective three-point function will be,
%%%%%%%%%%%%%%%%%%%%%%%%%%%%%%%%%%%%%%%%%%%%%%%%%%%%%%%%%%%%%%%%%%%%%%%%
\begin{equation}
\frac{d S_1}{d \mu_q}={-S_1 \Gamma_0 S_1}.
\end{equation}
%%%%%%%%%%%%%%%%%%%%%%%%%%%%%%%%%%%%%%%%%%%%%%%%%%%%%%%%%%%%%%%%%%%%%%%%
Rearranging terms in (\ref{eq.dsigmadmu_def}) it is possible
to write $\frac{d \sigma}{d \mu_q}$ in a closed form in terms of $m_0$,
$T$, $\mu_q$ and $\sigma$ only:
%%%%%%%%%%%%%%%%%%%%%%%%%%%%%%%%%%%%%%%%%%%%%%%%%%%%%%%%%%%%%%%%%%%%%%%%
\begin{equation}
 \frac{d\sigma}{d\mu_q}=
 \frac{-i{\rm Tr}(S_1\gamma_0 S_1)}{1+iG{\rm Tr}({S_1}^2)}.
 \label{eq.dsigmadmu_final}
\end{equation}
%%%%%%%%%%%%%%%%%%%%%%%%%%%%%%%%%%%%%%%%%%%%%%%%%%%%%%%%%%%%%%%%%%%%%%%%
For the second order derivative, one may start from
Eq.(\ref{eq.dsigmadmu_def}) to get,
%%%%%%%%%%%%%%%%%%%%%%%%%%%%%%%%%%%%%%%%%%%%%%%%%%%%%%%%%%%%%%%%%%%%%%%%
\begin{equation}
 \frac{d^2\sigma}{d\mu_q^2}=
2i\mathrm {Tr}[S_1 \Gamma_0 S_1 \Gamma_0 S_1]-
i\mathrm{Tr}\Big(S_1 G\frac{d^2 \sigma}{d \mu_q^2}S_1\Big).
\label{eq.d2sigdmu2_def}
\end{equation}
%%%%%%%%%%%%%%%%%%%%%%%%%%%%%%%%%%%%%%%%%%%%%%%%%%%%%%%%%%%%%%%%%%%%%%%%
Again rearranging terms, one can write a closed form expression for
$\frac{d^2\sigma}{d\mu_q^2}$ as a function of $m_0$, $T$, $\mu_q$,
$\sigma$ and $\frac{d \sigma}{d \mu_q}$ as,
%%%%%%%%%%%%%%%%%%%%%%%%%%%%%%%%%%%%%%%%%%%%%%%%%%%%%%%%%%%%%%%%%%%%%%%%
\begin{eqnarray}
 \frac{d^2\sigma}{d\mu_q^2}=
\frac{2i{\rm Tr}(S_1\Gamma_0 S_1\Gamma_0 S_1)}{1+iG{\rm Tr}({S_1}^2)}
=\frac{2i[{\rm{Tr}}(S_1\gamma_0 S_1\gamma_0 S_1)
+2(G\frac{d \sigma}{d \mu_q}){\rm{Tr}}({S_1}^3\gamma_0)
+(G\frac{d \sigma}{d \mu_q})^2{\rm{Tr}}({S_1}^3)]}
{1+iG{\rm Tr}({S_1}^2)}.
\end{eqnarray}
%%%%%%%%%%%%%%%%%%%%%%%%%%%%%%%%%%%%%%%%%%%%%%%%%%%%%%%%%%%%%%%%%%%%%%%%
We have plotted the first and second order derivatives of $\sigma$
w.r.t.\ $\mu_q$ as function of $T$ in Fig.\ref{fg.fldsder1_njl}
and Fig.\ref{fg.fldsder2_njl} respectively, where we have compared
the results of semi-analytical approach presented here to that of
numerical methods like Taylor expansion or finite difference.
%%%%%%%%%%%%%%%%%%%%%%%%%%%%%%%%%%%%%%%%%%%%%%%%%%%%%%%%%%%%%%%%%%%%
%%%%%%%%%%%%%%%%%%%%%%%%%%%%%%%%%%%%%%%%%%%%%%%%%%%%%%%%%%%%%%%%%%%%%

In the same way all the higher order derivatives may be obtained 
systematically as a function of $m_0$, $T$, $\mu_q$ and derivatives up
to one lower order.

This method is certainly more accurate than a direct numerical
differentiation of $\sigma$ w.r.t.\ $\mu_q$, or fitting Taylor
coefficients in an expansion w.r.t.\ $\mu_q$. No numerical
approximations or uncertainties are introduced, except for
the numerical integration of the fermionic momentum integrals.
Therefore the question of insensitivity at higher orders also
does not arise. 

%---------------------------------------------------------------------%
\subsection{PNJL model}
%---------------------------------------------------------------------%
We now discuss the Polyakov loop enhanced NJL model. The situation here
is similar to that of the NJL model except that we now have a couple of
mean fields more in the form of the expectation value of the Polyakov
loop $\Phi$ and that of its conjugate $\bar{\Phi}$. The Lagrangian for
the 2 flavor PNJL model is given by,
%%%%%%%%%%%%%%%%%%%%%%%%%%%%%%%%%%%%%%%%%%%%%%%%%%%%%%%%%%%%%%%%%%%%%%%%
\begin{equation}
 {\mathcal L}_{\rm PNJL} = \bar{\psi}(i\slashed D-m_0+\gamma_0\mu)\psi +
\frac{G}{2}[(\bar{\psi}\psi)^2+(\bar{\psi}i\gamma_5\vec{\tau}\psi)^2]
- {\mathcal U}(\Phi[A],\bar{\Phi}[A],T).
\end{equation}
%%%%%%%%%%%%%%%%%%%%%%%%%%%%%%%%%%%%%%%%%%%%%%%%%%%%%%%%%%%%%%%%%%%%%%%%
\noindent
where $D^\mu=\partial^\mu-ig{\mathcal A}^\mu_a\lambda_a/2$, 
${\mathcal A}^\mu_a$ being the $SU(3)$ background fields and $\lambda_a$
are the Gell-Mann matrices. Here the effective Polyakov loop potential is
given by,
%%%%%%%%%%%%%%%%%%%%%%%%%%%%%%%%%%%%%%%%%%%%%%%%%%%%%%%%%%%%%%%%%%%%%%%%
\begin{equation}
 \frac{{\mathcal U}(\Phi,\bar{\Phi},T)}{T^4} = 
    -\frac{b_2(T)}{2}\Phi\bar{\Phi} -
    \frac{b_3}{6}(\Phi^3+{\bar{\Phi}}^3) +
    \frac{b_4}{4}(\bar{\Phi}\Phi)^2,
\label{eq.potential}
\end{equation}
%%%%%%%%%%%%%%%%%%%%%%%%%%%%%%%%%%%%%%%%%%%%%%%%%%%%%%%%%%%%%%%%%%%%%%%%
with
%%%%%%%%%%%%%%%%%%%%%%%%%%%%%%%%%%%%%%%%%%%%%%%%%%%%%%%%%%%%%%%%%%%%%%%%
\begin{equation*}
 b_2(T) = a_0 + a_1\Big(\frac{T_0}{T}\Big) + a_2\Big(\frac{T_0}{T}\Big)^2 +
    a_3\Big(\frac{T_0}{T}\Big)^3.
\end{equation*}
%%%%%%%%%%%%%%%%%%%%%%%%%%%%%%%%%%%%%%%%%%%%%%%%%%%%%%%%%%%%%%%%%%%%%%%%
$\Phi$ is Polyakov loop and $\bar \Phi$ is its charge conjugate
\cite{rajarshidaB}. Values of coefficients $a_0,a_1,a_2,a_3,b_3,b_4$
have been taken from Ref.\cite{ratti1}. To take into account the effect
of SU(3) Haar measure in the PNJL model, we consider the modified
thermodynamic potential defined as \cite{rajarshidaC},
%%%%%%%%%%%%%%%%%%%%%%%%%%%%%%%%%%%%%%%%%%%%%%%%%%%%%%%%%%%%%%%%%%%%%%%%
\begin{equation}
 \Omega^\prime=\Omega-\kappa T^4 \ln[J(\Phi,{\bar \Phi})],
\end{equation}
%%%%%%%%%%%%%%%%%%%%%%%%%%%%%%%%%%%%%%%%%%%%%%%%%%%%%%%%%%%%%%%%%%%%%%%%
where $J(\Phi,{\bar \Phi})$ is the Vandermonde (VdM)
determinant given as,
%%%%%%%%%%%%%%%%%%%%%%%%%%%%%%%%%%%%%%%%%%%%%%%%%%%%%%%%%%%%%%%%%%%%%%%%
\begin{equation}
J[\Phi, {\bar \Phi}]=(27/24{\pi^2})(1-6\Phi {\bar \Phi}+
4(\Phi^3+{\bar \Phi}^3)-3{(\Phi {\bar \Phi})}^2).\nonumber\\
\end{equation}
%%%%%%%%%%%%%%%%%%%%%%%%%%%%%%%%%%%%%%%%%%%%%%%%%%%%%%%%%%%%%%%%%%%%%%%%
Following Ref.\cite{rajarshidaC}, the pressure in PNJL model is defined
as $\mathcal{P}=-\Omega$. 

%%%%%%%%%%%%%%%%%%%%%%%%%%%%%%%%%%%%%%%%%%%%%%%%%%%%%%%%%%%%%%%%%%%%
\begin{figure} [!ht]
\subfigure []
{\includegraphics [scale=0.6] {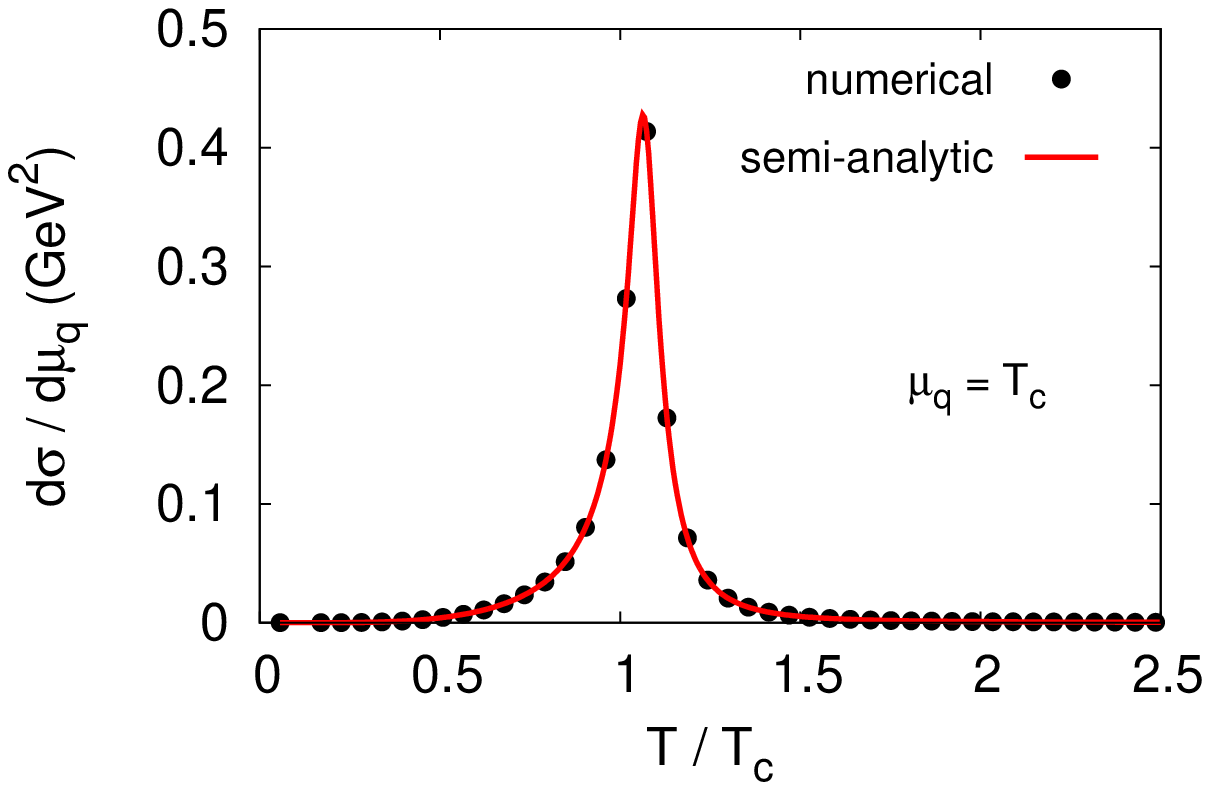}
\label{fg.sigder1_pnjl_nonzeromu}}
\subfigure []
{\includegraphics [scale=0.6] {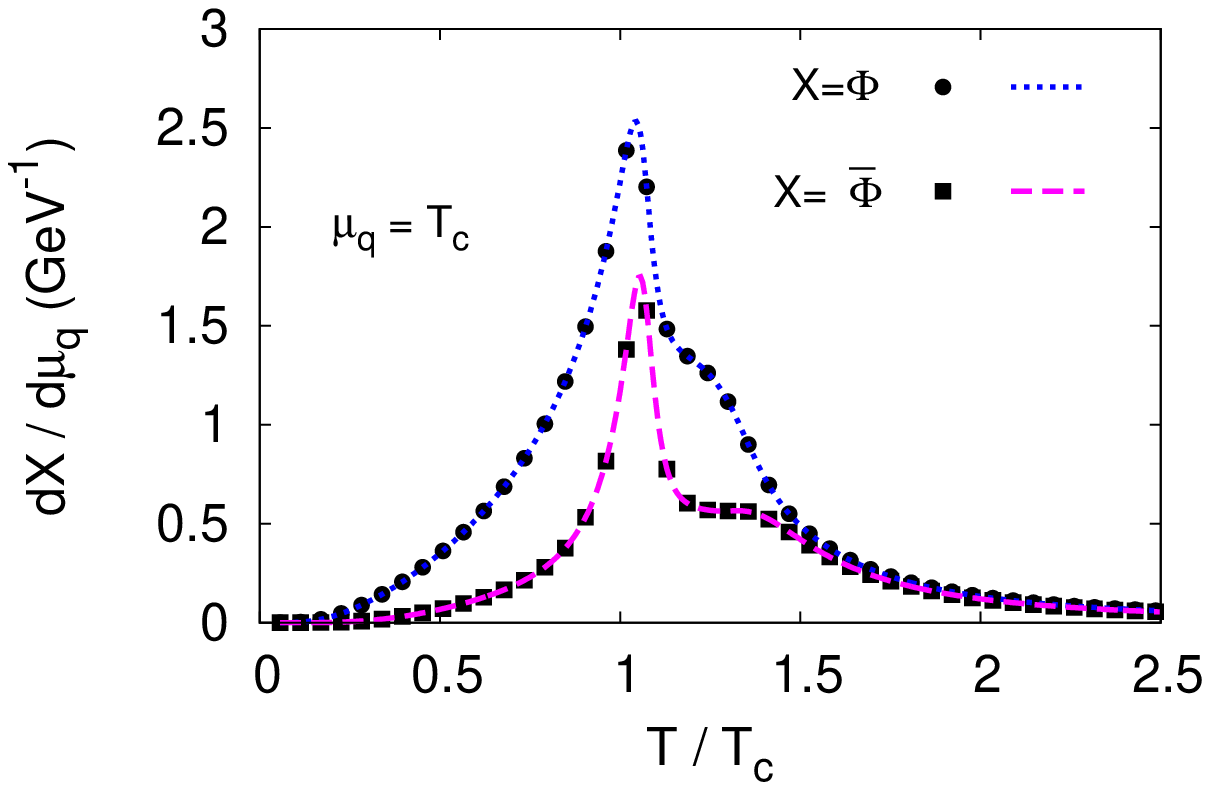}
\label{fg.phiphibder1_pnjl_nonzeromu}}
\caption{
(a) $\frac{d\sigma}{d\mu_q}$ and (b) $\frac{d\Phi}{d\mu_q}$ and
$\frac{d\bar{\Phi}}{d\mu_q}$ in the PNJL model at $\mu_q=T_c$. Here
the points represent the result from numerical differentiation, and
lines are from the semi-analytical approach described in the text.
}
\label{fg.fldsder_pnjl}
\end{figure}
%%%%%%%%%%%%%%%%%%%%%%%%%%%%%%%%%%%%%%%%%%%%%%%%%%%%%%%%%%%%%%%%%%%%%
In NJL model we obtained $\frac{d\sigma}{d\mu_q}$ by differentiating
the gap equation (\ref{eq.sig_def}) w.r.t.\ $\mu_q$. In the PNJL
model we do not have such gap equations for the $\Phi$ and $\bar{\Phi}$
fields. We therefore differentiate the stationarity conditions 
$\frac{\partial\Omega'}{\partial X}=0$ for $X=\Phi,\bar{\Phi},\sigma$
directly with the $\mu_q$ derivatives to get,
%%%%%%%%%%%%%%%%%%%%%%%%%%%%%%%%%%%%%%%%%%%%%%%%%%%%%%%%%%%%%%%%%%%%
\begin{equation}
\frac{d}{d\mu_q}\Big(\frac{\partial\Omega'}{\partial X}\Big)=0.
\label{eq.firstdermuq}
\end{equation}
%%%%%%%%%%%%%%%%%%%%%%%%%%%%%%%%%%%%%%%%%%%%%%%%%%%%%%%%%%%%%%%%%%%%
Note that this equation is valid only if we insert the mean field
values in $\frac{\partial\Omega'}{\partial X}$ before taking the
$\mu_q$ derivatives. This immediately gives us,
%%%%%%%%%%%%%%%%%%%%%%%%%%%%%%%%%%%%%%%%%%%%%%%%%%%%%%%%%%%%%%%%%%%%
\begin{eqnarray}
\frac{\partial}{\partial\mu_q}\Big(\frac{\partial\Omega'}{\partial X}\Big)+
\frac{\partial}{\partial\Phi}\Big(\frac{\partial\Omega'}{\partial X}\Big)
\cdot\frac{d\Phi}{d\mu_q}+
\frac{\partial}{\partial\bar{\Phi}}\Big(\frac{\partial\Omega'}{\partial X}\Big)
\cdot\frac{d\bar{\Phi}}{d\mu_q}+
\frac{\partial}{\partial\sigma}\Big(\frac{\partial\Omega'}{\partial X}\Big)
\cdot\frac{d\sigma}{d\mu_q}=0.
\end{eqnarray}
%%%%%%%%%%%%%%%%%%%%%%%%%%%%%%%%%%%%%%%%%%%%%%%%%%%%%%%%%%%%%%%%%%%%
So we have the matrix equation of the form $\bf {A\cdot x=B}$,
where $\bf {A}$ is the
coefficient matrix of the variables ${\bf x}=(\frac{d\Phi}{d\mu_q},
\frac{d\bar{\Phi}}{d\mu_q},\frac{d\sigma}{d\mu_q})^{\rm T}$ and
$\bf B$ matrix has the form ${\bf B}=(-\frac{\partial}{\partial\mu_q}
(\frac{\partial\Omega'}{\partial \Phi}),
-\frac{\partial}{\partial\mu_q}(\frac{\partial\Omega'}{\partial \bar{\Phi}}),
-\frac{\partial}{\partial\mu_q}(\frac{\partial\Omega'}{\partial \sigma}))^{\rm T}$.
The above matrix equation has the solutions of the form;
%%%%%%%%%%%%%%%%%%%%%%%%%%%%%%%%%%%%%%%%%%%%%%%%%%%%%%%%%%%%%%%%%%%%
\begin{equation}
\frac{d\Phi}{d\mu_q}=\frac{\triangle_1}{\triangle},
~~~~
\frac{d\bar{\Phi}}{d\mu_q}=\frac{\triangle_2}{\triangle},
~~~~
\frac{d\sigma}{d\mu_q}=\frac{\triangle_3}{\triangle},
\end{equation}
%%%%%%%%%%%%%%%%%%%%%%%%%%%%%%%%%%%%%%%%%%%%%%%%%%%%%%%%%%%%%%%%%%%%
where the Cramer's determinants are given by,
%%%%%%%%%%%%%%%%%%%%%%%%%%%%%%%%%%%%%%%%%%%%%%%%%%%%%%%%%%%%%%%%%%%%
\begin{equation}
\triangle = \rm{det}(\bf A) =
\left|
\begin{array}{ccc}
\dfrac{\partial}{\partial\Phi}\Big(\dfrac{\partial\Omega'}{\partial\Phi}\Big) &
\dfrac{\partial}{\partial\bar{\Phi}}\Big(\dfrac{\partial\Omega'}{\partial\Phi}\Big) &
\dfrac{\partial}{\partial\sigma}\Big(\dfrac{\partial\Omega'}{\partial\Phi}\Big)
\vspace{0.1in} \\
\dfrac{\partial}{\partial\Phi}\Big(\dfrac{\partial\Omega'}{\partial\bar{\Phi}}\Big) &
\dfrac{\partial}{\partial\bar{\Phi}}\Big(\dfrac{\partial\Omega'}{\partial\bar{\Phi}}\Big) &
\dfrac{\partial}{\partial\sigma}\Big(\dfrac{\partial\Omega'}{\partial\bar{\Phi}}\Big)
\vspace{0.1in} \\
\dfrac{\partial}{\partial\Phi}\Big(\dfrac{\partial\Omega'}{\partial\sigma}\Big) &
\dfrac{\partial}{\partial\bar{\Phi}}\Big(\dfrac{\partial\Omega'}{\partial\sigma}\Big) &
\dfrac{\partial}{\partial\sigma}\Big(\dfrac{\partial\Omega'}{\partial\sigma}\Big)
\end{array}
\right|,
\end{equation}
\vspace{0.2in}
\begin{equation}
\triangle_1 =
-\left|
\begin{array}{ccc}
\dfrac{\partial}{\partial\mu_q}\Big(\dfrac{\partial\Omega'}{\partial\Phi}\Big) &
\dfrac{\partial}{\partial\bar{\Phi}}\Big(\dfrac{\partial\Omega'}{\partial\Phi}\Big) &
\dfrac{\partial}{\partial\sigma}\Big(\dfrac{\partial\Omega'}{\partial\Phi}\Big)
\vspace{0.1in} \\
\dfrac{\partial}{\partial\mu_q}\Big(\dfrac{\partial\Omega'}{\partial\bar{\Phi}}\Big) &
\dfrac{\partial}{\partial\bar{\Phi}}\Big(\dfrac{\partial\Omega'}{\partial\bar{\Phi}}\Big) &
\dfrac{\partial}{\partial\sigma}\Big(\dfrac{\partial\Omega'}{\partial\bar{\Phi}}\Big)
\vspace{0.1in} \\
\dfrac{\partial}{\partial\mu_q}\Big(\dfrac{\partial\Omega'}{\partial\sigma}\Big) &
\dfrac{\partial}{\partial\bar{\Phi}}\Big(\dfrac{\partial\Omega'}{\partial\sigma}\Big) &
\dfrac{\partial}{\partial\sigma}\Big(\dfrac{\partial\Omega'}{\partial\sigma}\Big)
\end{array}
\right|,
\end{equation}
\vspace{0.2in}
\begin{equation}
\triangle_2 =
-\left|
\begin{array}{ccc}
\dfrac{\partial}{\partial\Phi}\Big(\dfrac{\partial\Omega'}{\partial\Phi}\Big) &
\dfrac{\partial}{\partial\mu_q}\Big(\dfrac{\partial\Omega'}{\partial\Phi}\Big) &
\dfrac{\partial}{\partial\sigma}\Big(\dfrac{\partial\Omega'}{\partial\Phi}\Big)
\vspace{0.1in} \\
\dfrac{\partial}{\partial\Phi}\Big(\dfrac{\partial\Omega'}{\partial\bar{\Phi}}\Big) &
\dfrac{\partial}{\partial\mu_q}\Big(\dfrac{\partial\Omega'}{\partial\bar{\Phi}}\Big) &
\dfrac{\partial}{\partial\sigma}\Big(\dfrac{\partial\Omega'}{\partial\bar{\Phi}}\Big)
\vspace{0.1in} \\
\dfrac{\partial}{\partial\Phi}\Big(\dfrac{\partial\Omega'}{\partial\sigma}\Big) &
\dfrac{\partial}{\partial\mu_q}\Big(\dfrac{\partial\Omega'}{\partial\sigma}\Big) &
\dfrac{\partial}{\partial\sigma}\Big(\dfrac{\partial\Omega'}{\partial\sigma}\Big)
\end{array}
\right|,
\end{equation}
\vspace{0.2in}
\begin{equation}
\triangle_3 =
-\left|
\begin{array}{cccc}
\dfrac{\partial}{\partial\Phi}\Big(\dfrac{\partial\Omega'}{\partial\Phi}\Big) &
\dfrac{\partial}{\partial\bar{\Phi}}\Big(\dfrac{\partial\Omega'}{\partial\Phi}\Big) &
\dfrac{\partial}{\partial\mu_q}\Big(\dfrac{\partial\Omega'}{\partial\Phi}\Big)
\vspace{0.1in} \\
\dfrac{\partial}{\partial\Phi}\Big(\dfrac{\partial\Omega'}{\partial\bar{\Phi}}\Big) &
\dfrac{\partial}{\partial\bar{\Phi}}\Big(\dfrac{\partial\Omega'}{\partial\bar{\Phi}}\Big) &
\dfrac{\partial}{\partial\mu_q}\Big(\dfrac{\partial\Omega'}{\partial\bar{\Phi}}\Big)
\vspace{0.1in} \\
\dfrac{\partial}{\partial\Phi}\Big(\dfrac{\partial\Omega'}{\partial\sigma}\Big) &
\dfrac{\partial}{\partial\bar{\Phi}}\Big(\dfrac{\partial\Omega'}{\partial\sigma}\Big) &
\dfrac{\partial}{\partial\mu_q}\Big(\dfrac{\partial\Omega'}{\partial\sigma}\Big)
\end{array}
\right|.
\end{equation}
%%%%%%%%%%%%%%%%%%%%%%%%%%%%%%%%%%%%%%%%%%%%%%%%%%%%%%%%%%%%%%%%%%%%

The elements of the determinants can be obtained from the expression
of $\Omega'$. Instead of using Cramer's rule, the solutions can also
be obtained through Gaussian Elimination method.
In Fig.\ref{fg.fldsder_pnjl} we have plotted the first order
derivatives of the mean fields along $T$ and compared the
results from two different methods as in case of NJL
in Fig.\ref{fg.fldsder_njl}.

One may similarly obtain the higher derivatives of the mean fields
with respect to $\mu_q$ by sequentially increasing the order of 
derivatives to act upon (\ref{eq.firstdermuq}). The derivatives at
any order will depend upon the various thermodynamic parameters as
well as the derivatives up to one lower order. For these higher
orders also we shall have to solve the similar matrix equations of
the form $\bf A \cdot x=B$. The most interesting part is that while
the column matrix $\bf B$ changes for every order, the coefficient
matrix $\bf A$ will remain same as in first order. We expect that
this semi-analytical prescription for obtaining field derivatives
will certainly give better results specifically in higher order,
compared to numerical derivatives. The detailed study in
that direction will be presented elsewhere. Here we shall use this
elegant formulation to gain some insight into the physics aspects
of susceptibilities in the context of fluctuation-dissipation theorem.

%=====================================================================%
\section{QNS from FDT}
\label{sc.qns_fdt}
%---------------------------------------------------------------------%
\subsection{A Toy Model}
\label{sc.qns_njl}
%%%%%%%%%%%%%%%%%%%%%%%%%%%%%%%%%%%%%%%%%%%%%%%%%%%%%%%%%%%%%%%%%%%%%%%%
\begin{figure} [!ht]
{\includegraphics [scale=0.4] {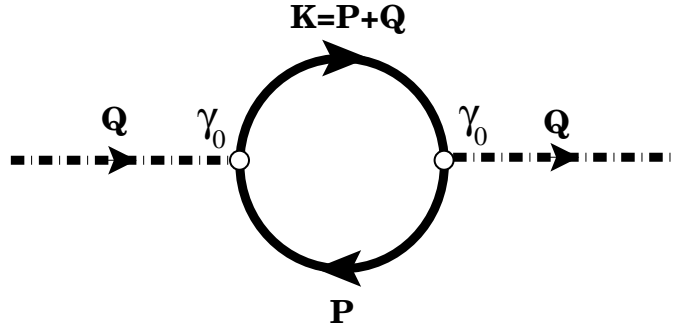}}
\caption{Time-time component of vector correlator.}
\label{fg.corr_free}
\end{figure}
%%%%%%%%%%%%%%%%%%%%%%%%%%%%%%%%%%%%%%%%%%%%%%%%%%%%%%%%%%%%%%%%%%%%%%%%
We consider a Lagrangian with an effective mass 
%%%%%%%%%%%%%%%%%%%%%%%%%%%%%%%%%%%%%%%%%%%%%%%%%%%%%%%%%%%%%%%%%%%%%%%%
\begin{equation}
\mathcal{L}=\bar{\psi}(i\slashed \partial
-\hat{M} +\gamma_0 \hat{\mu})\psi 
\label{eq.lag_mass}.
\end{equation}
%%%%%%%%%%%%%%%%%%%%%%%%%%%%%%%%%%%%%%%%%%%%%%%%%%%%%%%%%%%%%%%%%%%%%%%%
where $\hat{M}$ is effective mass matrix and $\hat{\mu}$ is the
matrix of chemical potential and both are diagonal in flavor space.
$\gamma_0$ is the three point function
\footnote{Equivalent to a massive free theory.} and
the corresponding two-point function for flavor $f$ is
%%%%%%%%%%%%%%%%%%%%%%%%%%%%%%%%%%%%%%%%%%%%%%%%%%%%%%%%%%%%%%%%%%%%%%%%
\begin{equation*}
S_f(L)=\frac{1}{\slashed L-M_f+{\mu}_f\gamma_0},
\end{equation*}
%%%%%%%%%%%%%%%%%%%%%%%%%%%%%%%%%%%%%%%%%%%%%%%%%%%%%%%%%%%%%%%%%%%%%%%%
where $L$ is the four momentum.
With this simple consideration, it is obvious that
Fig.\ref{fg.corr_free} is the relevant diagram in one loop that would
contribute to the time-time component of vector correlator $\Pi_{00}$,
%%%%%%%%%%%%%%%%%%%%%%%%%%%%%%%%%%%%%%%%%%%%%%%%%%%%%%%%%%%%%%%%%%%%%%%%
\begin{equation}
\Pi_{00}(q_0=\omega,q=\lvert\vec{Q}\rvert)=
-i\sum_{f=u,d}\int{\frac{d^4P}{(2\pi)^4}{\mathrm {Tr}}_{D,c}
[\gamma_0 S_f(K)\gamma_0 S_f(P)]},
\label{eq.pidef}
\end{equation}
%%%%%%%%%%%%%%%%%%%%%%%%%%%%%%%%%%%%%%%%%%%%%%%%%%%%%%%%%%%%%%%%%%%%%%%%
with $K=P+Q$. Here trace $\mathrm {Tr}_{D,c}$ is over Dirac and
color indices only. Replacing
\[\int \frac{dP_0}{2\pi}\longrightarrow \frac{i}{\beta} \sum_{\omega_n}~~,\]
and performing the Dirac trace, Eq.(\ref{eq.pidef}) becomes,
%%%%%%%%%%%%%%%%%%%%%%%%%%%%%%%%%%%%%%%%%%%%%%%%%%%%%%%%%%%%%%%%%%%%%%%%
\begin{eqnarray}
\Pi_{00}(\omega,q)&=& \sum_{f=u,d}\sum_n \frac{4}{\beta} 
\int\frac{d^3p}{(2\pi)^3}{\mathrm{Tr}}_c
\Big\{ \frac{(i\omega_n+\omega+\tilde{\mu}_f)
(i\omega_n+\tilde{\mu}_f)+\vec{P}\cdot\vec{K}+M_f^2}
{[(i\omega_n+\omega+\tilde{\mu}_f)^2-E_{fk}^2]
[(i\omega_n+\tilde{\mu}_f)^2-E_{fp}^2]} \Big\}.
\end{eqnarray}
%%%%%%%%%%%%%%%%%%%%%%%%%%%%%%%%%%%%%%%%%%%%%%%%%%%%%%%%%%%%%%%%%%%%%%%%
Now the remaining trace operation is over color space only.
Breaking into partial fractions R.H.S.\ of last equation can
be written as \cite{Klevansky:rmp},
%%%%%%%%%%%%%%%%%%%%%%%%%%%%%%%%%%%%%%%%%%%%%%%%%%%%%%%%%%%%%%%%%%%%%%%%
\begin{eqnarray}
\Pi_{00}(\omega,q) &=& \sum_{f=u,d}\sum_n \frac{1}{\beta}
\int\frac{d^3p}{(2\pi)^3}\frac{1}{E_{fp}E_{fk}}
{\mathrm{Tr}}_c \nonumber \\
&& \Big\{
\frac{1}{(i\omega_n+\omega+\tilde{\mu}_f)-E_{fk}}
\frac{1}{(i\omega_n+\tilde{\mu}_f)-E_{fp}}
[E_{fp}E_{fk}+M_f^2+\vec{P}\cdot\vec{K}] \nonumber \\
&+& \frac{1}{(i\omega_n+\omega+\tilde{\mu}_f)-E_{fk}}\frac{1}
{(i\omega_n+\tilde{\mu}_f)+E_{fp}}
[E_{fp}E_{fk}-M_f^2-\vec{P}\cdot\vec{K}] \nonumber \\
&+& \frac{1}{(i\omega_n+\omega+\tilde{\mu}_f)+E_{fk}}\frac{1}
{(i\omega_n+\tilde{\mu}_f)-E_{fp}}
[E_{fp}E_{fk}-M_f^2-\vec{P}\cdot\vec{K}] \nonumber \\
&+& \frac{1}{(i\omega_n+\omega+\tilde{\mu}_f)+E_{fk}}\frac{1}
{(i\omega_n+\tilde{\mu}_f)+E_{fp}}
[E_{fp}E_{fk}+M_f^2+\vec{P}\cdot\vec{K}]
\Big\}.
\end{eqnarray}
%%%%%%%%%%%%%%%%%%%%%%%%%%%%%%%%%%%%%%%%%%%%%%%%%%%%%%%%%%%%%%%%%%%%%%%%
Now performing the Matsubara summation over the discrete
frequencies, $\omega_n=(2n+1)\pi T$, we are left with,
%%%%%%%%%%%%%%%%%%%%%%%%%%%%%%%%%%%%%%%%%%%%%%%%%%%%%%%%%%%%%%%%%%%%%%%%
\begin{eqnarray}
\Pi_{00}(\omega,q) &=& \sum_{f=u,d}
\int\frac{d^3p}{(2\pi)^3}\frac{1}{E_{fp}E_{fk}}
{\mathrm{Tr}}_c \nonumber \\
&& \Big\{
\frac{E_{fp}E_{fk}+M_f^2+\vec{P}\cdot\vec{K}}{\omega+E_{fp}-E_{fk}}
[\mathcal{F}_1(E_{fp}-\tilde{\mu}_f)-
\mathcal{F}_1(E_{fk}-\tilde{\mu}_f-\omega)] \nonumber \\
&+&\frac{E_{fp}E_{fk}-M_f^2-\vec{P}\cdot\vec{K}}{\omega-E_{fp}-E_{fk}}
[1-\mathcal{F}_1(E_{fp}+\tilde{\mu}_f)-
\mathcal{F}_1(E_{fk}-\tilde{\mu}_f-\omega)] \nonumber \\
&+&\frac{E_{fp}E_{fk}-M_f^2-\vec{P}\cdot\vec{K}}{\omega+E_{fp}+E_{fk}}
[\mathcal{F}_1(E_{fp}-\tilde{\mu}_f)-1+
\mathcal{F}_1(E_{fk}+\tilde{\mu}_f+\omega)] \nonumber \\
&+&\frac{E_{fp}E_{fk}+M_f^2+\vec{P}\cdot\vec{K}}{\omega-E_{fp}+E_{fk}}
[-\mathcal{F}_1(E_{fp}+\tilde{\mu}_f)+
\mathcal{F}_1(E_{fk}+\tilde{\mu}_f+\omega)]
\Big\},
\label{eq.pi_after_sum}
\end{eqnarray}
%%%%%%%%%%%%%%%%%%%%%%%%%%%%%%%%%%%%%%%%%%%%%%%%%%%%%%%%%%%%%%%%%%%%%%%%
where ${\cal F}$ is the Fermi-Dirac distribution function.
If we make a change of variable $\vec{P}\rightarrow -\vec{P'}-\vec{Q}$,
in the third and fourth term then dot product of 3-vectors remains
unchanged and the momentum label of quasiparticle energy just
interchanges. Moreover keeping in mind that $e^{\beta \omega}=1$,
after simplification Eq.(\ref{eq.pi_after_sum}) becomes,
%%%%%%%%%%%%%%%%%%%%%%%%%%%%%%%%%%%%%%%%%%%%%%%%%%%%%%%%%%%%%%%%%%%%%%%%
\begin{eqnarray}
\Pi_{00}(\omega,q) &=& \sum_{f=u,d}
\int\frac{d^3p}{(2\pi)^3}\frac{1}{E_{fp}E_{fk}} \nonumber \\
&& {\mathrm{Tr}}_c
\Big\{
\frac{E_{fp}E_{fk}+M_f^2+\vec{P}\cdot\vec{K}}{\omega+E_{fp}-E_{fk}}
[\mathcal{F}(E_{fp}-{\mu}_f)+\mathcal{F}(E_{fp}+{\mu}_f)
-\mathcal{F}(E_{fk}-{\mu}_f)-
\mathcal{F}(E_{fk}+{\mu}_f)] \nonumber \\
&+& [E_{fp}E_{fk}-M_f^2-\vec{P}\cdot\vec{K}]
\Big[\frac{1}{\omega-E_{fp}-E_{fk}}-
\frac{1}{\omega+E_{fp}+E_{fk}}\Big][1-\mathcal{F}(E_{fp}+{\mu}_f)
-\mathcal{F}(E_{fk}-{\mu}_f)]
\Big\}.
\label{eq.total_pi} \nonumber \\
\end{eqnarray}
%%%%%%%%%%%%%%%%%%%%%%%%%%%%%%%%%%%%%%%%%%%%%%%%%%%%%%%%%%%%%%%%%%%%%%%%

%%%%%%%%%%%%%%%%%%%%%%%%%%%%%%%%%%%%%%%%%%%%%%%%%%%%%%%%%%%%%%%%%%%%%%%%
\subsubsection{Calculation of $\chi_q$ from real part of $\Pi_{00}$}
%----------------------------------------------------------------------%
After taking the real part of $\Pi_{00}(\omega,q)$ when we put $\omega=0$,
we are left with;
%%%%%%%%%%%%%%%%%%%%%%%%%%%%%%%%%%%%%%%%%%%%%%%%%%%%%%%%%%%%%%%%%%%%%%%%
\begin{eqnarray}
{\textrm {Re}} \Pi_{00}(\omega=0,q) &=& \sum_{f=u,d}
\int\frac{d^3p}{(2\pi)^3}\frac{1}{E_{fp}E_{fk}}{\mathrm{Tr}}_c 
\{
[E_{fp}E_{fk}+M_f^2+\vec{P}\cdot\vec{K}] \nonumber \\
&& \times\frac{\mathcal{F}(E_{fp}-{\mu}_f)+
\mathcal{F}(E_{fp}
+{\mu}_f)-\mathcal{F}(E_{fk}-{\mu}_f)-
\mathcal{F}(E_{fk}+{\mu}_f)}
{E_{fp}-E_{fk}} \nonumber \\
&-& 2
\frac{E_{fp}E_{fk}-M_f^2-\vec{P}\cdot\vec{K}}{E_{fp}+E_{fk}}
[1-\mathcal{F}(E_{fp}+{\mu}_f)
-\mathcal{F}(E_{fk}-{\mu}_f)]
\}.
\label{eq.re_pi}
\end{eqnarray}
%%%%%%%%%%%%%%%%%%%%%%%%%%%%%%%%%%%%%%%%%%%%%%%%%%%%%%%%%%%%%%%%%%%%%%%%
Now we are going to use the FDT as in Eq.(\ref{eq.fluc_diss_def}).
In the limit $q\rightarrow 0$ the second term of (\ref{eq.re_pi})
vanishes and for the first term
taking care of the $\frac{0}{0}$ form using L'Hospital rule we get,
%%%%%%%%%%%%%%%%%%%%%%%%%%%%%%%%%%%%%%%%%%%%%%%%%%%%%%%%%%%%%%%%%%%%%%%%
\begin{eqnarray}
\chi_q = 2\beta \sum_{f=u,d} \int\frac{d^3p}{(2\pi)^3} {\mathrm{Tr}}_c
\Big\{
\frac{e^{\beta(E_{fp}-{\mu}_f)}}
{(1+e^{\beta(E_{fp}-{\mu}_f)})^2} +
\frac{e^{\beta(E_{fp}+{\mu}_f)}}
{(1+e^{\beta(E_{fp}+{\mu}_f)})^2}
\Big\}. \label{eq.re_qns}
\end{eqnarray}
%%%%%%%%%%%%%%%%%%%%%%%%%%%%%%%%%%%%%%%%%%%%%%%%%%%%%%%%%%%%%%%%%%%%%%%%

\subsubsection{Calculation of $\chi_q$ from Imaginary part of $\Pi_{00}$}
%----------------------------------------------------------------------%
The imaginary part of the retarded correlator can be calculated
from the discontinuity in the following way;
%%%%%%%%%%%%%%%%%%%%%%%%%%%%%%%%%%%%%%%%%%%%%%%%%%%%%%%%%%%%%%%%%%%%%%%%
\begin{eqnarray}
{\textrm {Im}} \Pi_{00}(\omega,q) &=& \frac{1}{2i}
[\Pi_{00}(\omega\rightarrow \omega +i\eta,q)
-\Pi_{00}(\omega\rightarrow \omega -i\eta,q)] \nonumber \\
&=& -\pi \sum_{f=u,d} \int\frac{d^3p}{(2\pi)^3} 
\frac{1}{E_{fp}E_{fk}} {\mathrm{Tr}}_c
\{ 
(E_{fp}E_{fk}+M_f^2+\vec{P}\cdot\vec{K}) \nonumber \\
&& [\mathcal{F}(E_{fp}-{\mu}_f)+
\mathcal{F}(E_{fp}+{\mu}_f)
-\mathcal{F}(E_{fk}-{\mu}_f)-
\mathcal{F}(E_{fk}+{\mu}_f)]
\delta(\omega+E_{fp}-E_{fk}) \nonumber \\
&+& [E_{fp}E_{fk}-M_f^2-\vec{P}\cdot\vec{K}]
[1-\mathcal{F}(E_{fp}+{\mu}_f)-
\mathcal{F}(E_{fk}-{\mu}_f)] \nonumber \\
&& [\delta(\omega-E_{fp}-E_{fk})-\delta(\omega+E_{fp}+E_{fk})]
\}.
\label{eq.im_pi}
\end{eqnarray}
%%%%%%%%%%%%%%%%%%%%%%%%%%%%%%%%%%%%%%%%%%%%%%%%%%%%%%%%%%%%%%%%%%%%%%%%
The delta function in the first term of R.H.S.\ of the above equation
represents the contribution from the scattering process and the first
delta function of the second term represents the pair creation process
for $\omega>0$ \cite{Gert_calc,weldon_prd28}. The prefactors containing
Fermi-Dirac distributions to both of the above-mentioned terms can be
rearranged to show that they basically account for the statistical
weights of corresponding processes. Similarly for $\omega<0$,
one can realize some processes \cite{weldon_prd28} corresponding
to the second delta function in the second term. Although
that will clearly violates energy conservation for $\omega>0$,
since quasiparticle energies are always positive and therefore
hereinafter this term will be dropped.

As an intermediate but important step we want to show that, the
first term of R.H.S.\ of the above equation can be written as
\cite{Laine_deltaomega},
%%%%%%%%%%%%%%%%%%%%%%%%%%%%%%%%%%%%%%%%%%%%%%%%%%%%%%%%%%%%%%%%%%%%%%%%
\begin{eqnarray}
&& (E_{fp}(\omega+E_{fp})+M_f^2+\vec{P}\cdot\vec{K})
\delta(\omega+E_{fp}-E_{fk}) \times \nonumber \\
&& [\mathcal{F}(E_{fp}-{\mu}_f)+
\mathcal{F}(E_{fp}+{\mu}_f)
-\mathcal{F}(E_{fp}-{\mu}_f+\omega)-
\mathcal{F}(E_{fp}+{\mu}_f+\omega)] \nonumber \\
&& \stackrel{q \rightarrow 0}{\approx}
-(E_{fp}(\omega+E_{fp})+M_f^2+\vec{P}^2) ~ \omega\delta(\omega)
~ [\mathcal{F'}(E_{fp}-{\mu}_f)+
\mathcal{F'}(E_{fp}+{\mu}_f)] \nonumber \\
&=& (E_{fp}(\omega+E_{fp})+M_f^2+\vec{P}^2)
~ \omega\delta(\omega) \times \nonumber \\
&& \Big[\mathcal{F}(E_{fp}-{\mu}_f)
\Big(1-\mathcal{F}(E_{fp}-{\mu}_f)\Big)+
\mathcal{F}(E_{fp}+{\mu}_f)
\Big(1-\mathcal{F}(E_{fp}+{\mu}_f)\Big)\Big].
\end{eqnarray}
%%%%%%%%%%%%%%%%%%%%%%%%%%%%%%%%%%%%%%%%%%%%%%%%%%%%%%%%%%%%%%%%%%%%%%%%
Here, for the time being we have omitted the integration and
prefactors. Proportionality of this term to $\omega\delta(\omega)$
is important because it is related to number conservation
\cite{purnendu1,purnendu2,najmul,Kunihiro_PLB,Fujii04,Gert_calc}.
Apart from that, this kind of zero mode contribution in the
spectral functions is significant and gives rise to a
constant contribution in finite temperature Euclidean correlator
\cite{Umeda,Karsh,munshi,Ding,sourendu_meson_fluc}.

From the FDT as in Eq.(\ref{eq.fluc_diss_def}) and
using (\ref{eq.im_pi}) we get,
%%%%%%%%%%%%%%%%%%%%%%%%%%%%%%%%%%%%%%%%%%%%%%%%%%%%%%%%%%%%%%%%%%%%%%%%
\begin{eqnarray}
\chi_q &=& \lim_{q\rightarrow 0}\beta \sum_{f=u,d} 
\int\frac{d^3p}{(2\pi)^3} \frac{1}{E_{fp}E_{fk}} {\mathrm{Tr}}_c 
\Big\{ 
(E_{fp}E_{fk}+M_f^2+\vec{P}\cdot\vec{K}) \nonumber \\
&& \frac{\mathcal{F}(E_{fp}-{\mu}_f)+
\mathcal{F}(E_{fp}+{\mu}_f)
-\mathcal{F}(E_{fk}-{\mu}_f)-
\mathcal{F}(E_{fk}+{\mu}_f)}
{1-e^{\beta(E_{fp}-E_{fk})}} \nonumber \\
&+& (E_{fp}E_{fk}-M_f^2-\vec{P}\cdot\vec{K})
[1-\mathcal{F}(E_{fp}+{\mu}_f)-
\mathcal{F}(E_{fk}-{\mu}_f)]
\frac{1}{1-e^{-\beta(E_{fp}+E_{fk})}}
\Big\}. \nonumber
\end{eqnarray}
%%%%%%%%%%%%%%%%%%%%%%%%%%%%%%%%%%%%%%%%%%%%%%%%%%%%%%%%%%%%%%%%%%%%%%%%
For the limit of vanishing external momentum second term vanishes.
Using L'Hospital rule for the first term we are left with,
%%%%%%%%%%%%%%%%%%%%%%%%%%%%%%%%%%%%%%%%%%%%%%%%%%%%%%%%%%%%%%%%%%%%%%%%
\begin{eqnarray}
\chi_q = 2\beta \sum_{f=u,d} \int\frac{d^3p}{(2\pi)^3} {\mathrm{Tr}}_c
\Big\{
\frac{e^{\beta(E_{fp}-{\mu}_f)}}
{(1+e^{\beta(E_{fp}-{\mu}_f)})^2} +
\frac{e^{\beta(E_{fp}+{\mu}_f)}}
{(1+e^{\beta(E_{fp}+{\mu}_f)})^2}
\Big\}.
\label{eq.qns_before_trace}
\end{eqnarray}
%%%%%%%%%%%%%%%%%%%%%%%%%%%%%%%%%%%%%%%%%%%%%%%%%%%%%%%%%%%%%%%%%%%%%%%%

\subsubsection{Calculation of $\chi_q$ from Thermodynamic derivative}
One can write the partition function corresponding to
(\ref{eq.lag_mass}) as
%%%%%%%%%%%%%%%%%%%%%%%%%%%%%%%%%%%%%%%%%%%%%%%%%%%%%%%%%%%%%%%%%%%%%%
\begin{eqnarray}
{\cal Z}(\beta,\{{\mu}_f\})=\int {\cal D}[\bar \psi] 
{\cal D}[{\psi}] 
e^{-i\int d^4x\mathcal{L}(\psi,{\bar \psi};\{{\mu}_f\})},
\label{eq.i3}
\end{eqnarray}
%%%%%%%%%%%%%%%%%%%%%%%%%%%%%%%%%%%%%%%%%%%%%%%%%%%%%%%%%%%%%%%%%%%%%%%%%
where $\beta$ is the inverse temperature. 
The pressure can be written as
%%%%%%%%%%%%%%%%%%%%%%%%%%%%%%%%%%%%%%%%%%%%%%%%%%%%%%%%%%%%%%%%%%%%%%%%%
\begin{equation}
{\cal P}=\frac{1}{\cal V}
\ln{\cal Z}(\beta, \{{\mu}_f\})
\ , \label{eq.i3_1}
\end{equation}
%%%%%%%%%%%%%%%%%%%%%%%%%%%%%%%%%%%%%%%%%%%%%%%%%%%%%%%%%%%%%%%%%%%%%%%%%
where the four-volume, ${\cal V}=\beta V$ with 
$V$ is the three-volume. One can straight away compute the pressure and
show that the $\chi_q$ obtained from it through thermodynamic derivative
with respect to ${\mu}_f$ is exactly the same as those obtained in
(\ref{eq.re_qns}) and (\ref{eq.qns_before_trace}). 

Nevertheless,  we now demonstrate this in a very general perspective. 
${\mathcal P}'$ can be obtained from (\ref{eq.i3_1}) as
%%%%%%%%%%%%%%%%%%%%%%%%%%%%%%%%%%%%%%%%%%%%%%%%%%%%%%%%%%%%%%%%%%%%%%%%%
\begin{eqnarray}
\frac{\partial{\mathcal P}}{\partial {\mu}_f}
&=&\frac{-i}{{\cal V}{\cal Z}}\ {\int {\cal D}[\bar \psi] 
{\cal D}[\psi] 
\int d^4x \ {\bar \psi}^f(x) \gamma_0 \psi^f(x)} \,
e^{ {-i\int d^4x {\cal L} (\psi,{\bar{\psi}};\{{\mu}_f\})}} \ .
\label{eq.i3_2} 
\end{eqnarray}
%%%%%%%%%%%%%%%%%%%%%%%%%%%%%%%%%%%%%%%%%%%%%%%%%%%%%%%%%%%%%%%%%%%%%%%%%
The quark propagator in a hot and dense medium is defined as
%%%%%%%%%%%%%%%%%%%%%%%%%%%%%%%%%%%%%%%%%%%%%%%%%%%%%%%%%%%%%%%%%%%%%%%%%
\begin{eqnarray}
S^f_{\alpha\sigma}(x,x')&=&\frac{
\int {\cal D}[\bar \psi] {\cal D}[\psi] 
\psi^f_{\alpha}(x){\bar \psi}^f_\sigma(x') \exp \left ({-i\int d^4x {\cal L}
(\psi,{\bar{\psi}}; \{{\mu}_f\})}\right)}
{ \int {\cal D}[\bar \psi] {\cal D}[\psi]
 \exp \left ({-i\int d^4x {\cal L}
(\psi,{\bar{\psi}};\{{\mu}_f\})}\right)} \ .
\label{eq.i3_3}
\end{eqnarray}
%%%%%%%%%%%%%%%%%%%%%%%%%%%%%%%%%%%%%%%%%%%%%%%%%%%%%%%%%%%%%%%%%%%%%%%%%
From Eq.(\ref{eq.i3_2}) one can write the quark number density as
%%%%%%%%%%%%%%%%%%%%%%%%%%%%%%%%%%%%%%%%%%%%%%%%%%%%%%%%%%%%%%%%%%%%%%%%%
\begin{eqnarray}
n_q=\sum_f n_f=\sum_f\frac{\partial{\cal P}}{\partial {\mu}_f}
&=& -i\sum_f \int\! \frac{d^4P}{(2\pi)^4} 
{\mbox{Tr}}_{D,c}\left [S_f(P) \gamma_0 \right ].
\label{eq.i3_4}
\end{eqnarray}
%%%%%%%%%%%%%%%%%%%%%%%%%%%%%%%%%%%%%%%%%%%%%%%%%%%%%%%%%%%%%%%%%%%%%%%%%
Likewise, one can also obtain QNS as 
%%%%%%%%%%%%%%%%%%%%%%%%%%%%%%%%%%%%%%%%%%%%%%%%%%%%%%%%%%%%%%%%%%%%%%%%%
\begin{eqnarray}
\chi_q=\sum_f \frac{\partial^2{\cal P}}{\partial {\mu}^2_f}
&=& i\sum_f \int\! \frac{d^4P}{(2\pi)^4} 
{\mbox{Tr}}_{D,c}\left [S_f(P) \gamma_0 S_f(P) \gamma_0\right ]
=-i\mathrm {Tr}[S(P) \gamma_0 S(P) \gamma_0]
\label{eq.i3_5}
\end{eqnarray}
%%%%%%%%%%%%%%%%%%%%%%%%%%%%%%%%%%%%%%%%%%%%%%%%%%%%%%%%%%%%%%%%%%%%%%%%%
where the relation in (\ref{eq.unit}) is used. Now one can see that the
(\ref{eq.i3_5}) corresponds to the temporal correlator in
Fig.\ref{fg.corr_free} but at the external momentum
$Q=(\omega_q,|{\vec q}|)=0$ or amputated external legs. This static
correlator has been computed as $C_{00}$ in Appendix \ref{sc.trace},
which is exactly equal to those obtained in (\ref{eq.re_qns}) and
(\ref{eq.qns_before_trace}). This shows that the FDT vis-a-vis the
thermodynamic sum rule is satisfied in a toy model.
 
\subsection{NJL Model}

\begin{figure} [!ht]
\subfigure[]
{\includegraphics [scale=0.45] {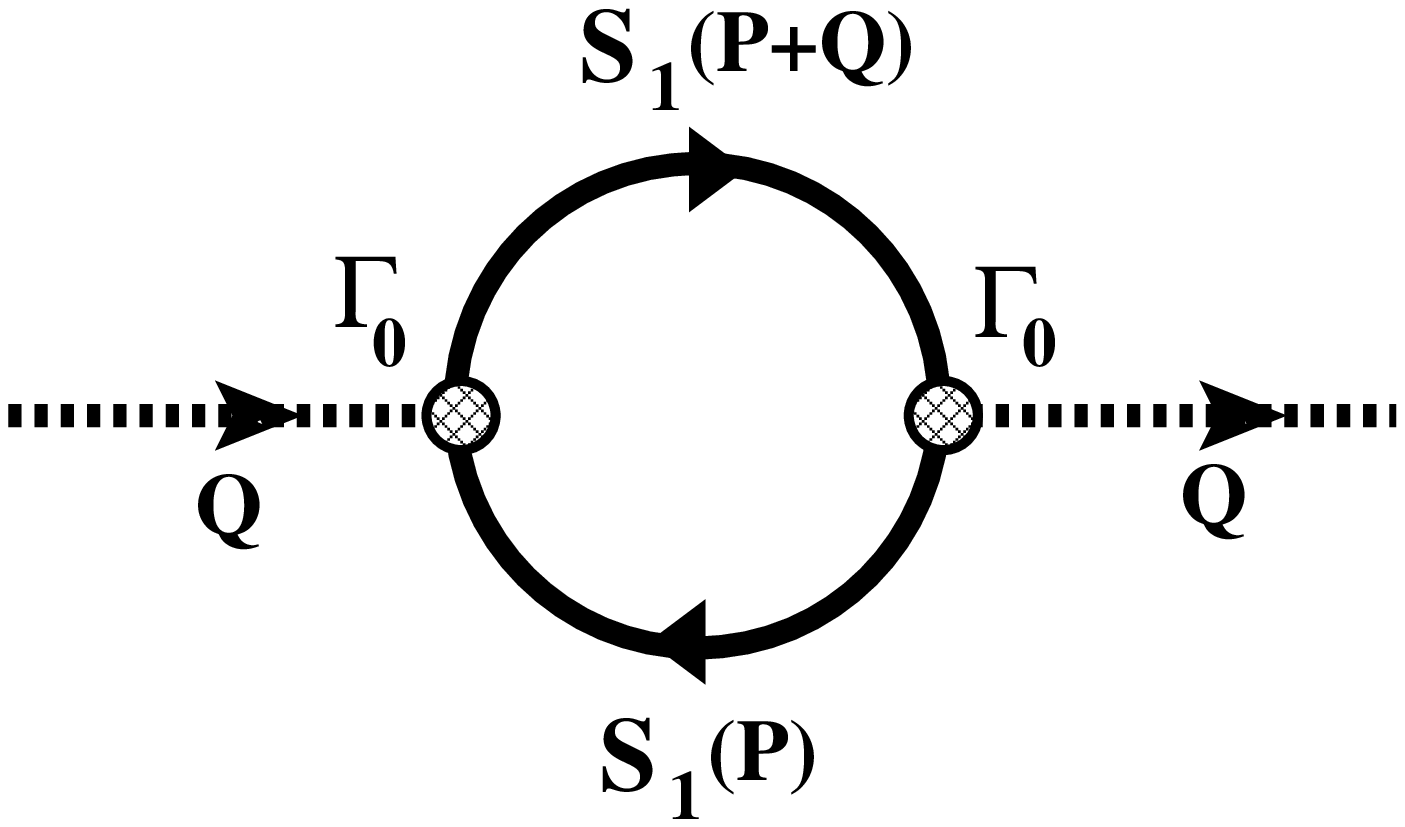}} ~~~~~~~~
\subfigure[]
{\includegraphics [scale=0.45] {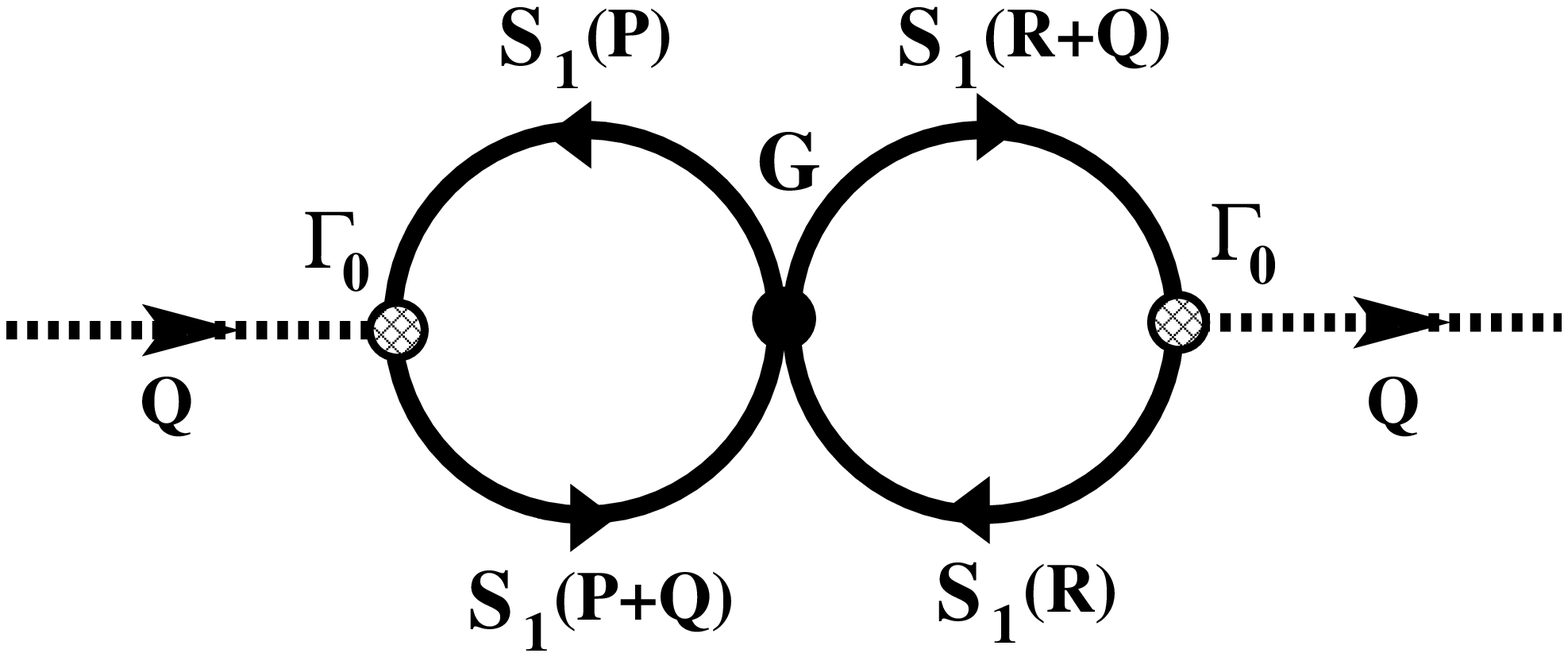}}
\caption{Time-time component of vector correlator in NJL model.
The diagram (a) is the usual self-energy with effective  propagator
and three-point vertex whereas diagram (b) has the origin of
four-fermionic interaction.}
\label{fg.corr_njl}
\end{figure}

So far our discussions are based on the naive consideration of a toy
model with a free massive propagator, but this has set the stage for
any realistic model calculations. Here, we consider the NJL
Lagrangian of Ref.(\ref{eq.lag_mf}), in which the explicit interaction
term through chiral condensate $\sigma$ is present and this would 
contribute to the physical quantities one would like to compute.
The relevant diagrams that would contribute to the correlation
function are shown in Fig.\ref{fg.corr_njl}.
%%%%%%%%%%%%%%%%%%%%%%%%%%%%%%%%%%%%%%%%%%%%%%%%%%%%%%%%%%%%%%%%%%%%%%%%%%
%%%%%%%%%%%%%%%%%%%%%%%%%%%%%%%%%%%%%%%%%%%%%%%%%%%%%%%%%%%%%%%%%%%
We note that the effective propagator $S_1$, three-point vertex
$\Gamma_0$ and the chiral condensates for NJL model have already
been defined in Sec.\ref{sc.fldsder_njl} and one can easily compute
these diagrams, but we purposefully avoid this and the reason for
which will be clear later.

Let us now first concentrate on the calculation of QNS from the
thermodynamic derivative of pressure with respect to the chemical
potential. As discussed earlier in Sec.\ref{sc.fldsder_njl}, the
mean fields have implicit dependences on chemical potential, thus
the thermodynamic derivatives are to be considered appropriately. 
This implies that one needs to consider the total derivative of
pressure rather than the explicit one. 

With these considerations, we write the pressure
${\cal P} = -\Omega$ from (\ref{eq.omega}). The quark number density is
then given as,
%%%%%%%%%%%%%%%%%%%%%%%%%%%%%%%%%%%%%%%%%%%%%%%%%%%%%%%%%%%%%%%%%%%%%%%%
\begin{eqnarray}
n_q= \frac{d \cal P}{d \mu_q}
&=&\frac{\partial \cal P}{\partial \mu_q}+
\frac{\partial \cal P} {\partial \sigma}\cdot\frac{d \sigma}{d \mu_q}
%\label{eq.dpdmu_tot}\\
= i\mathrm {Tr}(\gamma_0 S_1)+G \frac{d \sigma}{d \mu_q} 
[i\mathrm {Tr}(S_1)-\sigma]
%\label{eq.dpdmu}\\
= i\mathrm {Tr}(\Gamma_0 S_1)
-G\sigma\frac{d \sigma}{d \mu_q}, 
\label{eq.dp_dmu}
\end{eqnarray}
%%%%%%%%%%%%%%%%%%%%%%%%%%%%%%%%%%%%%%%%%%%%%%%%%%%%%%%%%%%%%%%%%%%%%%%%
where we have used
%%%%%%%%%%%%%%%%%%%%%%%%%%%%%%%%%%%%%%%%%%%%%%%%%%%%%%%%%%%%%%%%%%%%%%%%
\begin{equation}
{\frac{\partial \cal P}{\partial \mu_q}}= 
i\mathrm {Tr}(\gamma_0 S_1),
\label{eq.delp_delmu}
\end{equation}
%%%%%%%%%%%%%%%%%%%%%%%%%%%%%%%%%%%%%%%%%%%%%%%%%%%%%%%%%%%%%%%%%%%%%%%%
and
%%%%%%%%%%%%%%%%%%%%%%%%%%%%%%%%%%%%%%%%%%%%%%%%%%%%%%%%%%%%%%%%%%%%%%%%
\begin{equation}
\frac{\partial \cal P}{\partial \sigma}=G[i\mathrm {Tr}(S_1)-\sigma].
\label{eq.delp_delsigma}
\end{equation}
%%%%%%%%%%%%%%%%%%%%%%%%%%%%%%%%%%%%%%%%%%%%%%%%%%%%%%%%%%%%%%%%%%%%%%%%
It is interesting to note that for number density, $n_q$, the partial
derivative alone gives the full contribution in the mean field theory
as $\frac{\partial P}{\partial \sigma}=0$ if one uses (\ref{eq.sig_def}) in 
(\ref{eq.delp_delsigma}). 

One can similarly obtain the second order derivative of pressure w.r.t.\
$\mu_q$ from (\ref{eq.dp_dmu}) to get the QNS as,
%%%%%%%%%%%%%%%%%%%%%%%%%%%%%%%%%%%%%%%%%%%%%%%%%%%%%%%%%%%%%%%%%%%%%%%%
\begin{eqnarray}
\chi_q= \frac{d^2 \cal P}{d \mu_q^2}
 &=&\dfrac{\partial^2 \cal P}
{\partial \mu_q^2}
+\Big[\frac{\partial}{\partial \mu_q}
\Big(\frac{\partial \cal P}{\partial \sigma}\Big)
+\frac{\partial}{\partial \sigma}
\Big(\frac{\partial \cal P}{\partial \mu_q}\Big)\Big]
\cdot\frac{d \sigma}{d \mu_q}
+\frac{\partial \cal P}{\partial \sigma}\cdot
\frac{d^2 \sigma}{d \mu_q^2}
+\frac{\partial^2 \cal P}{\partial \sigma^2}\cdot
\Big(\frac{d \sigma}{d \mu_q}\Big)^2 \nonumber \\
 &=& -i \mathrm {Tr}[\gamma_0 S_1 \gamma_0 S_1]-
2iG\Big(\frac{d \sigma}{d \mu_q}\Big)\mathrm {Tr}[S_1 \gamma_0 S_1]+
G \frac{d^2 \sigma}{d \mu_q^2}[i \mathrm {Tr}(S_1)-\sigma]+
G\Big(\frac{d \sigma}{d \mu_q}\Big)^2[-iG \mathrm {Tr}(S_1^2)-1] \nonumber \\
 &=& -i \mathrm {Tr}(\Gamma_0 S_1\Gamma_0 S_1)
 -G\Big(\frac{d \sigma}{d \mu_q}\Big)^2
 +G \frac{d^2 \sigma}{d \mu_q^2}[i \mathrm {Tr}(S_1)-\sigma],
\label{eq.d2pdmu2b}
\end{eqnarray}
%%%%%%%%%%%%%%%%%%%%%%%%%%%%%%%%%%%%%%%%%%%%%%%%%%%%%%%%%%%%%%%%%%%%%%%%
where various second order explicit derivatives 
are used  in terms of respective correlators
by using (\ref{eq.delp_delmu}) and
(\ref{eq.delp_delsigma}). These relations are noted below:
%%%%%%%%%%%%%%%%%%%%%%%%%%%%%%%%%%%%%%%%%%%%%%%%%%%%%%%%%%%%%%%%%%%%%%%%
\begin{eqnarray}
\frac{\partial^2 \cal P}{\partial \mu_q^2}
&=&-i \mathrm {Tr}[\gamma_0 S_1 \gamma_0 S_1]=-C_{00},
\label{eq.explmuq}\\
\frac{\partial}{\partial \mu_q}
\Big(\frac{\partial \cal P}{\partial \sigma}\Big)
&=&-iG \mathrm {Tr}[S_1 \gamma_0 S_1]=-C_{01},
\label{eq.cross_mu_sigma2} \\
\frac{\partial}{\partial \sigma}
\Big(\frac{\partial \cal P}{\partial \mu_q}\Big)
&=&-iG \mathrm {Tr}[S_1 \gamma_0 S_1]=-C_{10},
\label{eq.cross_mu_sigma1}\\
\frac{\partial^2 \cal P}{\partial \sigma^2}
&=&G [-iG \mathrm {Tr}(S_1^2)-1]=-C_{11}-G.
\label{eq.delp_delsig2}
\end{eqnarray}
%%%%%%%%%%%%%%%%%%%%%%%%%%%%%%%%%%%%%%%%%%%%%%%%%%%%%%%%%%%%%%%%%%%%%%%%
The detailed calculations of the correlators $C_{00}, C_{01}$ and
$C_{11}$ are presented in Appendix \ref{sc.trace}. Here we have used the
relation $\dfrac{\partial S_1}{\partial\sigma} =-G{S_1}^2$,
which can be easily obtained from (\ref{eq.mod_prop_inv}).

Furthermore, the last term of (\ref{eq.d2pdmu2b}) vanishes due to
(\ref{eq.sig_def}) and using the first equality of
(\ref{eq.dsigmadmu_def}) we can finally write,
%%%%%%%%%%%%%%%%%%%%%%%%%%%%%%%%%%%%%%%%%%%%%%%%%%%%%%%%%%%%%%%%%%%%%%%%
\begin{equation}
\chi_q= \frac{d^2 \cal P}{d \mu_q^2}=
-i \mathrm {Tr}(\Gamma_0 S_1\Gamma_0 S_1)-
G(-i\mathrm {Tr}[S_1 \Gamma_0 S_1])^2.
\label{eq.d2pdmu2_corrl}
\end{equation}
%%%%%%%%%%%%%%%%%%%%%%%%%%%%%%%%%%%%%%%%%%%%%%%%%%%%%%%%%%%%%%%%%%%%%%%%

%%%%%%%%%%%%%%%%%%%%%%%%%%%%%%%%%%%%%%%%%%%%%%%%%%%%%%%%%%%%%%%%%%%%%%%%
\begin{figure} [!ht]
\subfigure[]
{\includegraphics [scale=0.45] {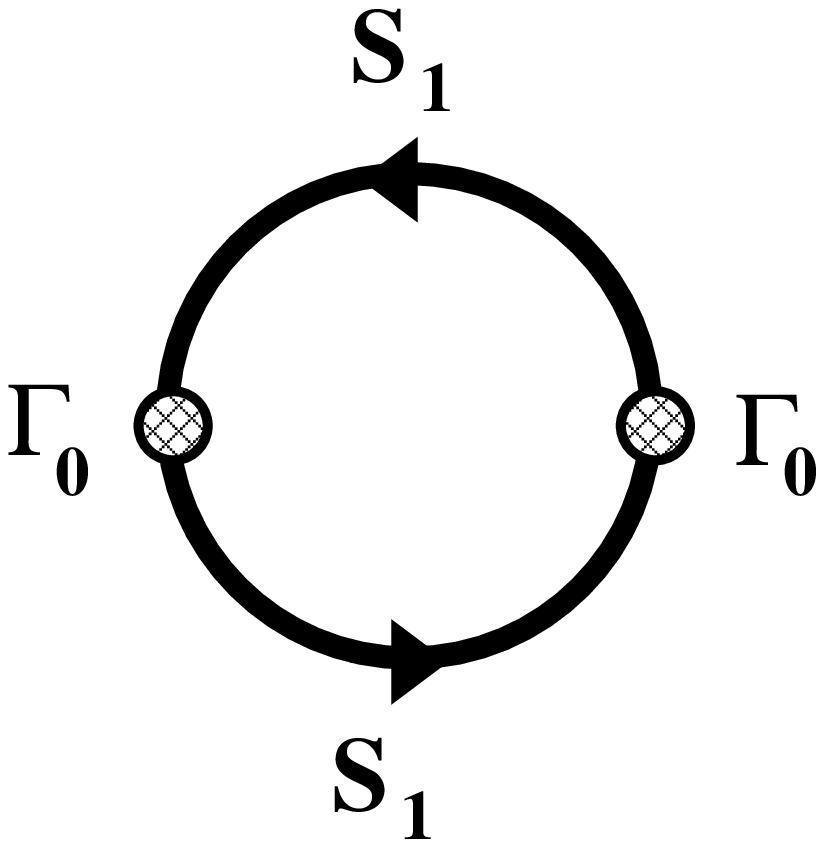}
\label{fg.corltr_con}} ~~~~~~~~
\subfigure[]
{\includegraphics [scale=0.45] {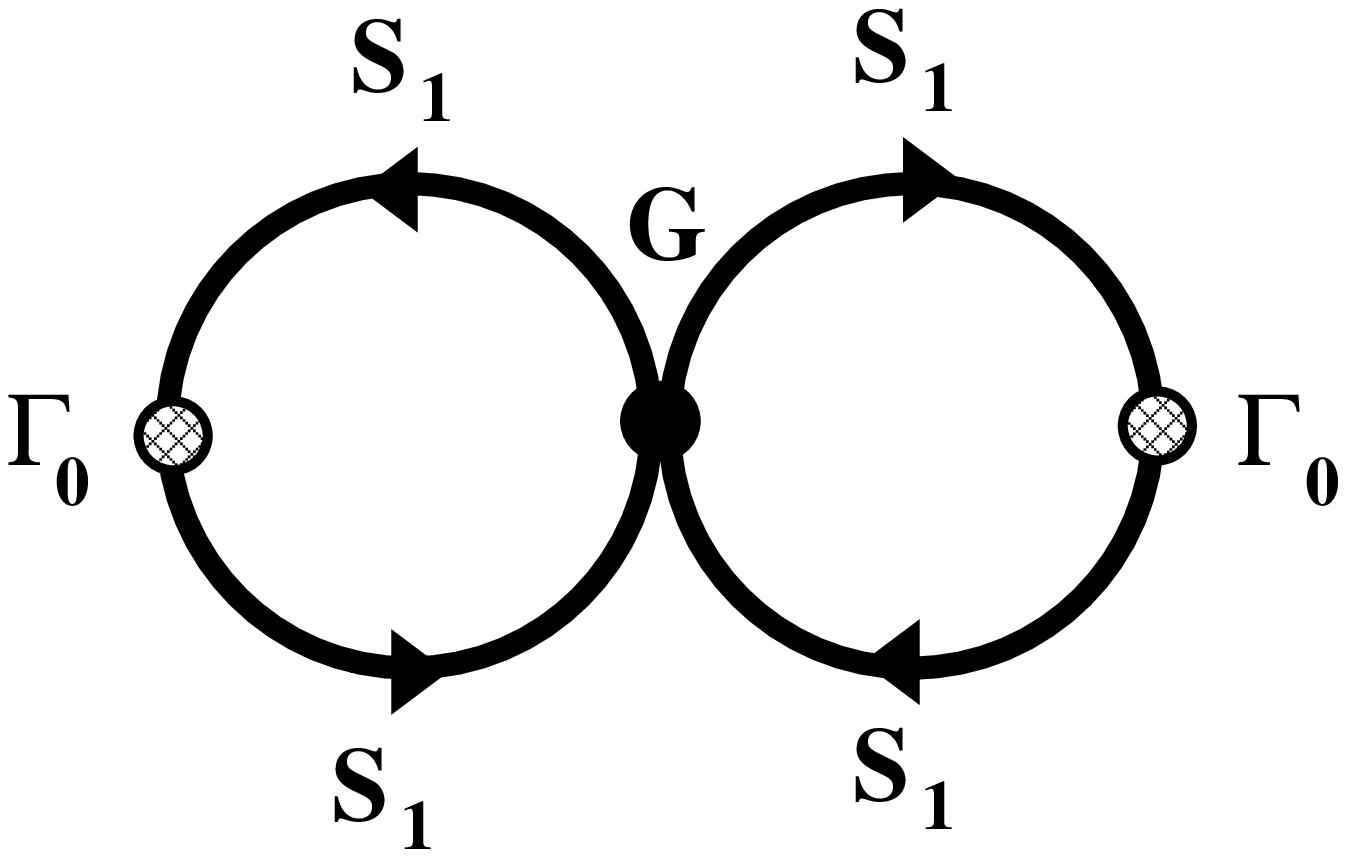}
\label{fg.corltr_dis}}
\caption{Static or amputated vector correlators with modified
propagator and effective three-point function.}
\label{fg.corltr_top}
\end{figure}
%%%%%%%%%%%%%%%%%%%%%%%%%%%%%%%%%%%%%%%%%%%%%%%%%%%%%%%%%%%%%%%%%%%%%%%%

The right hand side of (\ref{eq.d2pdmu2_corrl}) can be
viewed in terms of  diagrammatic topology as displayed
in Fig.\ref{fg.corltr_top}. It is evident that these are 
equivalent to the vector correlator in NJL model in static limit
or amputated legs as given in Fig.\ref{fg.corr_njl}. 
We note that in a mean field approach, where the mean fields 
are  sensitive to external source, an appropriate measure 
is to be taken to satisfy the FDT vis-a-vis thermodynamic sum
rule. This implies that the inclusion of 
implicit $\mu_q$ dependences of the mean fields is not ad hoc, 
rather it enables us, from the field theoretic point of view, 
to compute the correlators associated with the conserved density 
fluctuation through diagrammatic way in NJL model.

Another interesting as well as relevant point we would like to
demonstrate below. Putting the results of (\ref{eq.cross_mu_sigma2})
and (\ref{eq.delp_delsig2}) into (\ref{eq.dsigmadmu_final}),
we can rewrite it as
%%%%%%%%%%%%%%%%%%%%%%%%%%%%%%%%%%%%%%%%%%%%%%%%%%%%%%%%%%%%%%%%%%%%%%%%
\begin{equation}
\dfrac{d\sigma}{d\mu_q}=-\dfrac{\dfrac{\partial}{\partial \sigma}
\Big(\dfrac{\partial {\cal P}}{\partial \mu_q}\Big)}
{\dfrac{\partial^2 {\cal P}}{\partial \sigma^2}}
\label{eq.dsigmadmu_new}
\end{equation}
%%%%%%%%%%%%%%%%%%%%%%%%%%%%%%%%%%%%%%%%%%%%%%%%%%%%%%%%%%%%%%%%%%%%%%%%
Now  replacing this $\dfrac{d\sigma}{d\mu_q}$ in the first line of
Eq.(\ref{eq.d2pdmu2b}) and keeping in mind that the last but one term
will vanish due to the mean field condition, one can have 
%%%%%%%%%%%%%%%%%%%%%%%%%%%%%%%%%%%%%%%%%%%%%%%%%%%%%%%%%%%%%%%%%%%%%%%%
\begin{eqnarray}
\frac{d^2 {\cal P}}{d \mu_q^2}&=&\frac{\partial^2 P}{\partial \mu_q^2}
-2\frac{\Big[\frac{\partial}{\partial \sigma}
\Big(\frac{\partial {\cal P}}{\partial \mu_q}\Big)\Big]^2}
{\frac{\partial^2 {\cal P}}{\partial \sigma^2}}
+\frac{\partial^2 {\cal P}}{\partial \sigma^2}\cdot
\frac{\Big[\frac{\partial}{\partial \sigma}
\Big(\frac{\partial {\cal P}}{\partial \mu_q}\Big)\Big]^2}
{[\frac{\partial^2 {\cal P}}{\partial \sigma^2}]^2} \nonumber \\
&=&\frac{\frac{\partial^2 {\cal P}}{\partial \mu_q^2}\cdot
\frac{\partial^2 {\cal P}}{\partial \sigma^2}
-\Big[\frac{\partial}{\partial \sigma}
\Big(\frac{\partial {\cal P}}{\partial \mu_q}\Big)\Big]^2}
{\frac{\partial^2 {\cal P}}{\partial \sigma^2}} 
=\frac{\partial^2 {\cal P}}{\partial \mu_q^2}-
\frac{\partial^2 {\cal P}}{\partial\mu_q\partial\sigma}\cdot
\Big( \frac{\partial^2 {\cal P}}{\partial \sigma^2} \Big)^{-1}\cdot
\frac{\partial^2 {\cal P}}{\partial\sigma\partial\mu_q}
\end{eqnarray}
%%%%%%%%%%%%%%%%%%%%%%%%%%%%%%%%%%%%%%%%%%%%%%%%%%%%%%%%%%%%%%%%%%%%%%%%
This mixing pattern is already established in
Refs.\cite{Kunihiro_PLB,Fujii03} and  is similar to the
mixing in susceptibilities when computed from  the inverse of
curvature matrix \cite{Fukushima_PRD,Fukushima_PLB}.

%%%%%%%%%%%%%%%%%%%%%%%%%%%%%%%%%%%%%%%%%%%%%%%%%%%%%%%%%%%%%%%%%%%%%%%%
\begin{figure} [!ht]
{\includegraphics [scale=0.6] {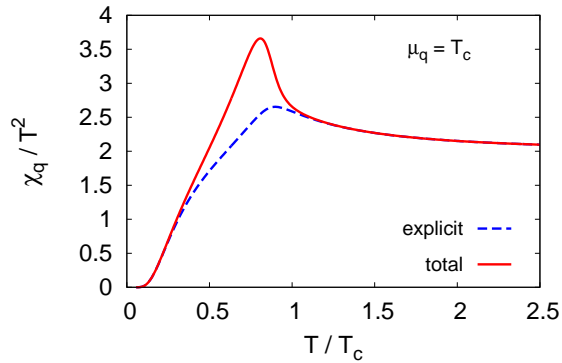}}
\caption{Quark number susceptibility in 2 flavor NJL model
at nonzero $\mu_q$.}
\label{fg.chi_q_njl}
\end{figure}
%%%%%%%%%%%%%%%%%%%%%%%%%%%%%%%%%%%%%%%%%%%%%%%%%%%%%%%%%%%%%%%%%%%%%%%%

The behavior of QNS in NJL model is shown in Fig.\ref{fg.chi_q_njl}.
Here the study has been done for 2 flavor NJL model at only at non-zero
$\mu_q$. This is because at $\mu_q=0$ we have
$\frac{d\sigma}{d\mu_q}=0$ due to CP symmetry, and thus all implicit
contributions vanish. The two curves in Fig.\ref{fg.chi_q_njl}
represent explicit and total contributions to the QNS respectively.
Important contributions from the implicit $\mu_q$ dependent terms
arise close to the transition region where the change in the mean fields
is most significant. It is needless to mention that the QNS obtained
here from (\ref{eq.d2pdmu2_corrl}) comes out to be the same as that
obtained from any of the numerical derivative methods.
%XXXXXXXXXXXXXXXXXXXXXXXXXXXXXXXXXXXXXXXXXXXXXXXXXXXXXXXXXXXXXXXXXXXXXX%

%----------------------------------------------------------------------%
\subsection{PNJL Model}
\label{sc.pnjl_qns}
%----------------------------------------------------------------------%
%%%%%%%%%%%%%%%%%%%%%%%%%%%%%%%%%%%%%%%%%%%%%%%%%%%%%%%%%%%%%%%%%%%%%%%%
\begin{figure} [!ht]
\subfigure[]
{\includegraphics [scale=0.6] {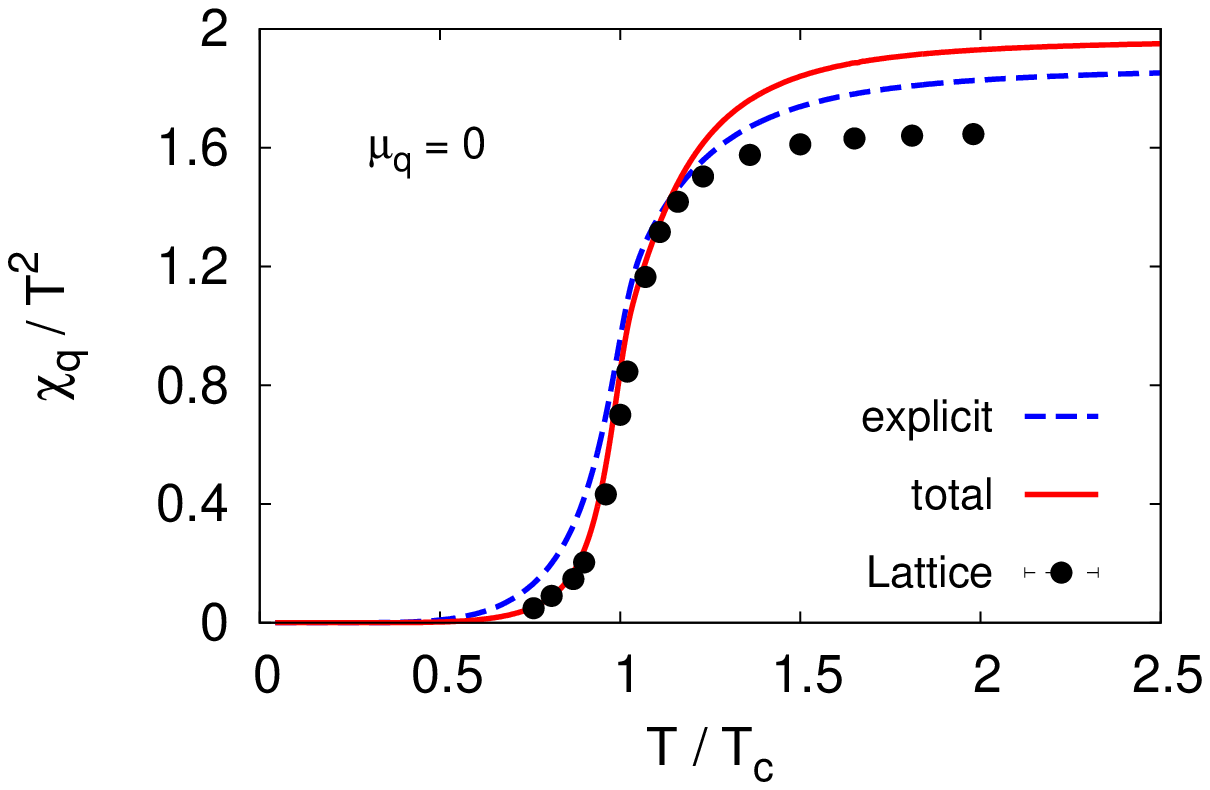}}
\subfigure[]
{\includegraphics [scale=0.6] {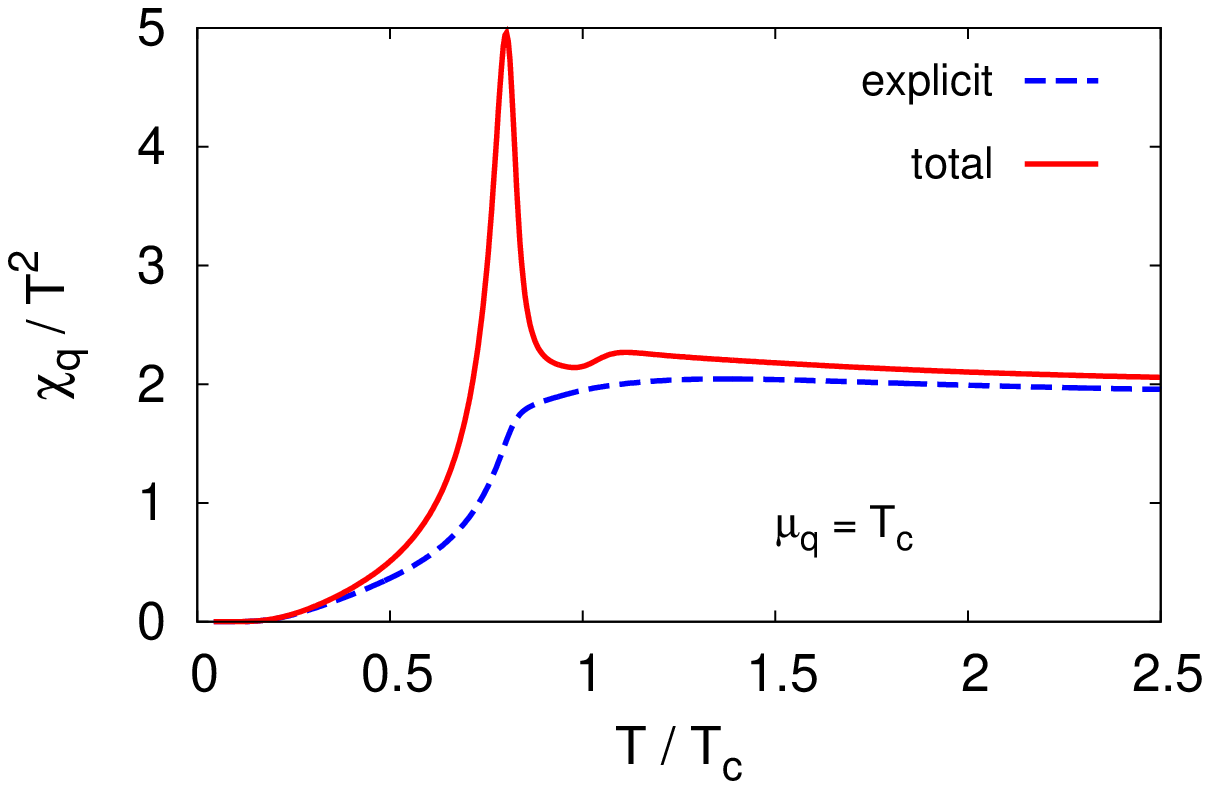}}
\caption{(a) Quark number susceptibility in 2 flavor PNJL model
at $\mu_q=0$. Lattice data are taken from Ref.\cite{Alton}.
(b) QNS in PNJL model for non-zero chemical potential.}
\label{fg.qns_nonzromu_pnjl}
\end{figure}
%%%%%%%%%%%%%%%%%%%%%%%%%%%%%%%%%%%%%%%%%%%%%%%%%%%%%%%%%%%%%%%%%%%%%%%%
In the present form of the PNJL model, the gluon physics is contained
only effectively in a static background field that comes through the
inclusion of Polyakov loop. We have no gluon-like quasi particles in
PNJL model. We treat such bosonic fields as purely classical ones unlike
the fermionic fields. One can only obtain their mean values through
the minimization of the thermodynamic potential. Obviously the Polyakov
loop fields are also to be determined from a set of transcendental
equations. The mean fields $\Phi$ and $\bar{\Phi}$ so obtained also
depend on $T$ and $\mu_q$ in a similar way as $\sigma$. Here one can
intuitively write as in the case of the NJL model that,
%%%%%%%%%%%%%%%%%%%%%%%%%%%%%%%%%%%%%%%%%%%%%%%%%%%%%%%%%%%%%%%%%%%%%%%%
\begin{equation}
\frac{d \cal P}{d \mu_q}=\frac{\partial \cal P}{\partial \mu_q}
+\frac{\partial \cal P}{\partial \sigma}\frac{d \sigma}{d \mu_q}
+\frac{\partial \cal P}{\partial \Phi}\frac{d \Phi}{d \mu_q}
+\frac{\partial \cal P}{\partial \bar{\Phi}}\frac{d \bar{\Phi}}{d \mu_q}
\equiv\frac{\partial \cal P}{\partial \mu_q}+\sum_{X=\sigma,\Phi,\bar{\Phi}}
\frac{\partial \cal P}{\partial X}\cdot\frac{dX}{d\mu_q}.
\label{eq.delp_delmu_pnjl}
\end{equation}
%%%%%%%%%%%%%%%%%%%%%%%%%%%%%%%%%%%%%%%%%%%%%%%%%%%%%%%%%%%%%%%%%%%%%%%%
Like in the NJL model, here also the second term of
(\ref{eq.delp_delmu_pnjl}) will vanish due to
$\frac{\partial \cal P}{\partial \sigma}=0$. But a crucial difference lies in
the fact that for the PNJL model,
$\frac{\partial \cal P}{\partial \Phi}\neq 0$ and
$\frac{\partial \cal P}{\partial \bar{\Phi}}\neq 0$.
This is due to the fact that $P=-\Omega\neq-\Omega'$ \cite{rajarshidaC}.
So, in case of the PNJL model even for first order derivative, we shall
have a finite contribution from implicit $\mu_q$ dependences through
$\Phi$ and $\bar{\Phi}$, i.e.
$\frac{d \cal P}{d \mu_q}\neq\frac{\partial \cal P}{\partial \mu_q}$.

Differentiating (\ref{eq.delp_delmu_pnjl}) w.r.t.\ $\mu_q$ we have,
%%%%%%%%%%%%%%%%%%%%%%%%%%%%%%%%%%%%%%%%%%%%%%%%%%%%%%%%%%%%%%%%%%%%%%%%
\begin{equation}
\frac{d^2 \cal P}{d\mu_q^2}=\frac{\partial^2 \cal P}{\partial\mu_q^2}+
2\sum_{X=\sigma,\Phi,\bar{\Phi}}
\frac{\partial^2 \cal P}{\partial\mu_q\partial X}\cdot
\frac{dX}{d\mu_q}+\sum_{X=\sigma,\Phi,\bar{\Phi}}
\frac{\partial \cal P}{\partial X}\cdot\frac{d^2X}{d\mu_q^2}
+\sum_{X,Y=\sigma,\Phi,\bar{\Phi}}
\frac{\partial^2 \cal P}{\partial X\partial Y}\cdot
\frac{dX}{d\mu_q}\cdot\frac{dY}{d\mu_q},
\label{eq.qns}                                 
\end{equation}
%%%%%%%%%%%%%%%%%%%%%%%%%%%%%%%%%%%%%%%%%%%%%%%%%%%%%%%%%%%%%%%%%%%%%%%%
where, the first term is from the explicit appearances of $\mu_q$
and the other three terms contains the contributions coming from
the implicit $\mu_q$ dependences of pressure through the mean fields.

In the left panel of  Fig.\ref{fg.qns_nonzromu_pnjl} the plots
of $\chi_q$ at $\mu_q=0$ are presented for the PNJL model. The
contribution from the explicit $\mu_q$ dependence and the total
contribution are shown separately. The latter contribution comes
out to be same as the QNS obtained from numerical derivatives of
pressure. Again due to CP symmetry, the non-vanishing implicit
contributions come through the $\mu_q$ dependence of $\Phi$ and
$\bar{\Phi}$ only. Our result is compared to that of
Lattice QCD\cite{Alton}. The QNS in the PNJL model for non-zero
$\mu_q$ is presented in the right panel of
Fig.\ref{fg.qns_nonzromu_pnjl}. Here again the explicit contribution
is shown separately. As is clearly evident, the presence of the
implicit contributions are significant close to the transition. The
most notable feature is that the peak in the susceptibility arises
solely due to the implicit chemical potential dependence of pressure.
Location of any critical point is therefore crucially dependent on the
proper evaluation of chemical potential dependence of the mean fields.
%%%%%%%%%%%%%%%%%%%%%%%%%%%%%%%%%%%%%%%%%%%%%%%%%%%%%%%%%%%%%%%%%

\section{Conclusion}
\label{sc.concl}
In QCD inspired model the fluctuation-dissipation theorem is
usually assumed to be applicable, and susceptibilities which are
associated with fluctuations are calculated from the derivatives
of pressure. In the present work through an extensive exercise we
have shown the fluctuation-dissipation theorem holds true within
the framework of NJL and PNJL models. In mean field approaches,
the mean fields are sensitive to the external source like quark
chemical potential, there should be additional contributions
coming from the implicit dependence of the mean fields on the
chemical potential. On the other hand, the temporal vector
correlator associated with the fluctuations is  modified due
to the effective interaction in these model Lagrangians. Here we
have given an elegant formalism and shown that the inclusion
of implicit dependent terms through the mean fields is actually
consistent with the field theoretic point of view and consolidates 
the fluctuation-dissipation theorem. For the NJL model a complete
analysis through diagrammatics could be found. While such elegant
exercise did not result for the PNJL model due to the classical
nature of the Polyakov loop, the essence of the modification
required has been clearly presented.

We have also described an analytical method for calculating the
derivative of the mean fields with respect to the chemical potential,
which forms the essential part of the modifications in the 
fluctuation-dissipation theorem. This approach is essential if one
intends to study higher order derivatives for which numerical 
differentiation becomes unreliable. Further studies in higher order
derivatives will be presented elsewhere. We expect these studies to
play an important role in understanding the nature of the critical
region in the phase diagram of strongly interacting matter.
%XXXXXXXXXXXXXXXXXXXXXXXXXXXXXXXXXXXXXXXXXXXXXXXXXXXXXXXXXXXXXXXXXXXXXX%

%======================================================================%
\begin{acknowledgments}
%----------------------------------------------------------------------%
We would like to thank Gert Aarts, Olaf Kaczmarek, Yan Zhu and
Mikko Laine for their crucial comments and useful communications.
Useful discussions with Najmul Haque and Chowdhury Aminul Islam are
also hereby acknowledged. A.L. acknowledges Rishi Sharma for
useful discussions.
\end{acknowledgments}
%XXXXXXXXXXXXXXXXXXXXXXXXXXXXXXXXXXXXXXXXXXXXXXXXXXXXXXXXXXXXXXXXXXXXXX%

%%%%%%%%%%%%%%%%%%%%%%%%%%%%%%%%%%%%%%%%%%%%%%%%%%%%%%%%%%%%%%%%%%%%%%%%
%%%%%% APPENDICES %%%%%%%%%%%%%%%%%%%%%%%%%%%%%%%%%%%%%%%%%%%%%%%%%%%%%%
%%%%%%%%%%%%%%%%%%%%%%%%%%%%%%%%%%%%%%%%%%%%%%%%%%%%%%%%%%%%%%%%%%%%%%%%
\appendix

%======================================================================%
\section{Calculation of traces in main text}
\label{sc.trace}
%----------------------------------------------------------------------%
As we are working in isospin symmetric limit, the current and
constituent masses of $u$ and $d$ flavors are equal. So, in all
the trace calculations of this section we will not carry the flavor
index of propagators and chemical potentials and whenever there is
trace or sum over flavors that will give only a factor of $N_f$.
%----------------------------------------------------------------------%
\subsection{Calculation of R.H.S.\ of Eq.(\ref{eq.explmuq})}
\label{sc.calc_chi}
\begin{figure} [!ht]
{\includegraphics [scale=0.35] {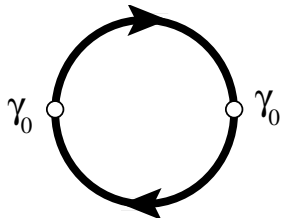}}
\caption{Time-time component of amputated vector correlator}
\label{fg.bare_correlator}
\end{figure}

%----------------------------------------------------------------------%
Here we discuss the calculation of the time-time component of the 
amputated correlator with bare vertices which is shown in
Fig.\ref{fg.bare_correlator} and discussed in the main text.
%%%%%%%%%%%%%%%%%%%%%%%%%%%%%%%%%%%%%%%%%%%%%%%%%%%%%%%%%%%%%%%%%%%%%
%%%%%%%%%%%%%%%%%%%%%%%%%%%%%%%%%%%%%%%%%%%%%%%%%%%%%%%%%%%%%%%%%%%%%
%%%%%%%%%%%%%%%%%%%%%%%%%%%%%%%%%%%%%%%%%%%%%%%%%%%%%%%%%%%%%%%%%%%%%
\begin{eqnarray}
C_{00} &=& i{\rm Tr}[\gamma_0 S_1(P)\gamma_0 S_1(P)]\nonumber\\
&=&-i\int{\frac{d^4P}{(2\pi)^4}{\rm Tr}_{D,f,c}
[\gamma_0 S_1(P)\gamma_0 S_1(P)]}.
\end{eqnarray}
%%%%%%%%%%%%%%%%%%%%%%%%%%%%%%%%%%%%%%%%%%%%%%%%%%%%%%%%%%%%%%%%%%%%%
with $S_1(P)$ being the quark propagator of momentum $P$
which is given by,
$S_1(P)=\frac{1}{\slashed P - M + \gamma_0\tilde{\mu}_q}$,
where, $\tilde{\mu}_q$ is the chemical potential and M is constituent
quark mass\footnote{For a toy model $M=m_0$ (say).} $M=m_0-G\sigma$.
For NJL $\tilde{\mu}_q=\mu_q$ and for PNJL $\tilde{\mu}_q=\mu_q-iA_4$.
In last equation \textquoteleft${\rm Tr}_{D,f,c}$\textquoteright\,
represents trace over Dirac, flavor and color indices only.
Making the replacement
%%%%%%%%%%%%%%%%%%%%%%%%%%%%%%%%%%%%%%%%%%%%%%%%%%%%%%%%%%%%%%%%%%%%%
\begin{equation*}
\int \frac{dP_0}{2\pi}\longrightarrow \frac{i}{\beta}
\sum_n~, ~~~~{\rm where~} {\omega_n=(2n+1)\frac{\pi}{\beta}},
\end{equation*}
%%%%%%%%%%%%%%%%%%%%%%%%%%%%%%%%%%%%%%%%%%%%%%%%%%%%%%%%%%%%%%%%%%%%%
%%%%%%%%%%%%%%%%%%%%%%%%%%%%%%%%%%%%%%%%%%%%%%%%%%%%%%%%%%%%%%%%%%%%%
\begin{eqnarray}
C_{00}&=&\frac{4N_f}{\beta}\sum_n\int\frac{d^3p}{(2\pi)^3}
\mathrm{Tr}_c{\frac{(i\omega_n+\tilde{\mu}_q)^2
+{E_p}^2}{[(i\omega_n+\tilde{\mu}_q)^2-{E_p}^2]^2}}\nonumber\\
&=&\frac{2N_f}{\beta}\sum_n\int\frac{d^3p}{(2\pi)^3}\mathrm{Tr}_c
\Big[\frac{1}{(i\omega_n+\tilde{\mu}_q-E_p)^2}
+\frac{1}{(i\omega_n+\tilde{\mu}_q+E_p)^2}\Big].
\end{eqnarray}
%%%%%%%%%%%%%%%%%%%%%%%%%%%%%%%%%%%%%%%%%%%%%%%%%%%%%%%%%%%%%%%%%%%%%
where, $E_p^2=p^2+M^2$. For Matsubara summation we use:
%%%%%%%%%%%%%%%%%%%%%%%%%%%%%%%%%%%%%%%%%%%%%%%%%%%%%%%%%%%%%%%%%%%%%
\begin{equation*}
\frac{1}{\beta}\sum_{n}\frac{1}{(i\omega_n\pm z)^2}=
-\frac{\beta e^{\beta z}}{(1+e^{\beta z})^2}.
\end{equation*}
%%%%%%%%%%%%%%%%%%%%%%%%%%%%%%%%%%%%%%%%%%%%%%%%%%%%%%%%%%%%%%%%%%%%%
So finally,
%%%%%%%%%%%%%%%%%%%%%%%%%%%%%%%%%%%%%%%%%%%%%%%%%%%%%%%%%%%%%%%%%%%%%
\begin{equation}
C_{00}=-2N_f\beta \int \frac{d^3p}{(2\pi)^3} \mathrm{Tr}_c
\Big[\frac{e^{\beta (E_p-\tilde{\mu}_q)}}
{(1+e^{\beta (E_p-\tilde{\mu}_q)})^2}+
\frac{e^{\beta (E_p+\tilde{\mu}_q)}}
{(1+e^{\beta (E_p+\tilde{\mu}_q)})^2}\Big].
\label{eq.correlator_final}
\end{equation}
%%%%%%%%%%%%%%%%%%%%%%%%%%%%%%%%%%%%%%%%%%%%%%%%%%%%%%%%%%%%%%%%%%%%%
For NJL model color trace is trivial, only the number of color $N_c$
will be factored out. We have,
%%%%%%%%%%%%%%%%%%%%%%%%%%%%%%%%%%%%%%%%%%%%%%%%%%%%%%%%%%%%%%%%%%%%%
\begin{equation}
C_{00}=-2N_fN_c\beta \int \frac{d^3p}{(2\pi)^3}
\Big[\frac{e^{\beta (E_p-\mu_q)}}
{(1+e^{\beta (E_p-\mu_q)})^2}+
\frac{e^{\beta (E_p+\mu_q)}}
{(1+e^{\beta (E_p+\mu_q)})^2}\Big].
\end{equation}
%%%%%%%%%%%%%%%%%%%%%%%%%%%%%%%%%%%%%%%%%%%%%%%%%%%%%%%%%%%%%%%%%%%%%
Now we are going to evaluate the color trace in PNJL model.
Let us assume, $\mathcal{F}_2(x)=\dfrac{e^x}{(1+e^x)^2}$.
Note the fact that,
%%%%%%%%%%%%%%%%%%%%%%%%%%%%%%%%%%%%%%%%%%%%%%%%%%%%%%%%%%%%%%%%%%%%%
\begin{equation*}
\frac{\partial^2}{\partial {E_p}^2} 
\ln [1+e^{-\beta(E_{p}-\tilde{\mu}_q)}]
= \beta^2 \mathcal{F}_2(E_{p}-\tilde{\mu}_q).
\end{equation*}
%%%%%%%%%%%%%%%%%%%%%%%%%%%%%%%%%%%%%%%%%%%%%%%%%%%%%%%%%%%%%%%%%%%%%
So, following Ref.\cite{ratti_meson} we can write,
%%%%%%%%%%%%%%%%%%%%%%%%%%%%%%%%%%%%%%%%%%%%%%%%%%%%%%%%%%%%%%%%%%%%%
\begin{eqnarray}
\mathrm{Tr}_c \mathcal{F}_2(E_{p}-\tilde{\mu}_q) &=& \frac{1}{\beta^2}
\frac{\partial^2}{\partial E_{p}^2} {\rm Tr}_c 
\ln [1+e^{-\beta(E_{p}-\tilde{\mu}_q)}] \nonumber \\
&=& \frac{1}{\beta^2} \frac{\partial^2}{\partial E_{p}^2} 
{\rm Tr}_c \ln [1+L^\dagger e^{-\beta(E_{p}-{\mu}_q)}] \nonumber \\
&=& 3 \Big[
\frac{\bar\Phi e^{-\beta(E_{p}-\mu_q)}
+4\Phi e^{-2\beta(E_{p}-\mu_q)}+3e^{-3\beta(E_{p}-\mu_q)}}
{1+3\bar\Phi e^{-\beta(E_{p}-\mu_q)}
+3\Phi e^{-2\beta(E_{p}-\mu_q)}+ e^{-3\beta(E_{p}-\mu_q)}} \nonumber \\
&-& 3 \frac{(\bar\Phi e^{-\beta(E_{p}-\mu_q)}
+2\Phi e^{-2\beta(E_{p}-\mu_q)}+e^{-3\beta(E_{p}-\mu_q)})^2}
{(1+3\bar\Phi e^{-\beta(E_{p}-\mu_q)}
+3\Phi e^{-2\beta(E_{p}-\mu_q)}+ e^{-3\beta(E_{p}-\mu_q)})^2}
\Big].
\label{eq.tr_final}
\end{eqnarray}
%%%%%%%%%%%%%%%%%%%%%%%%%%%%%%%%%%%%%%%%%%%%%%%%%%%%%%%%%%%%%%%%%%%%%
Similar expression can be found for
$\mathrm{Tr}_c \mathcal{F}_2(E_{p}+\mu_q)$.
Finally putting altogether these expressions we got,
%%%%%%%%%%%%%%%%%%%%%%%%%%%%%%%%%%%%%%%%%%%%%%%%%%%%%%%%%%%%%%%%%%%%%
\begin{eqnarray}
C_{00} &=& -6 N_f\beta \int\frac{d^3p}{(2\pi)^3}
\Big[
\frac{\bar\Phi e^{-\beta(E_{p}-\mu_q)}
+4\Phi e^{-2\beta(E_{p}-\mu_q)}+3e^{-3\beta(E_{p}-\mu_q)}}
{1+3\bar\Phi e^{-\beta(E_{p}-\mu_q)}
+3\Phi e^{-2\beta(E_{p}-\mu_q)}+ e^{-3\beta(E_{p}-\mu_q)}} 
\nonumber \\
&-& 3 \frac{(\bar\Phi e^{-\beta(E_{p}-\mu_q)}
+2\Phi e^{-2\beta(E_{p}-\mu_q)}+e^{-3\beta(E_{p}-\mu_q)})^2}
{(1+3\bar\Phi e^{-\beta(E_{p}-\mu_q)}
+3\Phi e^{-2\beta(E_{p}-\mu_q)}+ e^{-3\beta(E_{p}-\mu_q)})^2} 
\nonumber \\
&+& \frac{\Phi e^{-\beta(E_{p}+\mu_q)}
+4\bar\Phi e^{-2\beta(E_{p}+\mu_q)}+3e^{-3\beta(E_{p}+\mu_q)}}
{1+3\Phi e^{-\beta(E_{p}+\mu_q)}
+3\bar\Phi e^{-2\beta(E_{p}+\mu_q)}+ e^{-3\beta(E_{p}+\mu_q)}} 
\nonumber \\
&-& 3 \frac{(\Phi e^{-\beta(E_{p}+\mu_q)}
+2\bar\Phi e^{-2\beta(E_{p}+\mu_q)}+e^{-3\beta(E_{p}+\mu_q)})^2}
{(1+3\Phi e^{-\beta(E_{p}+\mu_q)}
+3\bar\Phi e^{-2\beta(E_{p}+\mu_q)}+ e^{-3\beta(E_{p}+\mu_q)})^2}
\Big].
\end{eqnarray}
%%%%%%%%%%%%%%%%%%%%%%%%%%%%%%%%%%%%%%%%%%%%%%%%%%%%%%%%%%%%%%%%%%%%%
%XXXXXXXXXXXXXXXXXXXXXXXXXXXXXXXXXXXXXXXXXXXXXXXXXXXXXXXXXXXXXXXXXXXXXX%

%----------------------------------------------------------------------%
\subsection{Calculation of R.H.S.\ of Eq.(\ref{eq.cross_mu_sigma1})
and Eq.(\ref{eq.cross_mu_sigma2})}
\label{sc.calc_cross}
%----------------------------------------------------------------------%
\begin{eqnarray}
C_{10}=C_{01}&=&iG{\rm Tr}
[S_1(P)\gamma_0 S_1(P)] \nonumber \\
&=&-iG\int{\frac{d^4P}{(2\pi)^4}{\rm Tr}_{D,f,c}
[S_1(P)\gamma_0 S_1(P)]} \nonumber \\
&=& \frac{4N_fG}{\beta}\sum_n\int\frac{d^3p}{(2\pi)^3}
\mathrm{Tr}_c\frac{2M(i\omega_n+\tilde{\mu}_q)}
{[(i\omega_n+\tilde{\mu}_q)^2-{E_p}^2]^2} \nonumber \\
&=& \frac{2N_f}{\beta}\sum_n\int\frac{d^3p}{(2\pi)^3}
\Big(\frac{GM}{E_p} \Big)\mathrm{Tr}_c
\Big[ \frac{1}{(i\omega_n+\tilde{\mu}_q-E_p)^2}-\frac{1}
{(i\omega_n+\tilde{\mu}_q+E_p)^2}  \Big]\\
&=&-2N_f\beta\int\frac{d^3p}{(2\pi)^3}
\Big(\frac{GM}{E_p} \Big)\mathrm{Tr}_c
 \Big[\frac{e^{\beta (E_p-\tilde{\mu}_q)}}
{(1+e^{\beta (E_p-\tilde{\mu}_q)})^2}-
\frac{e^{\beta (E_p+\tilde{\mu}_q)}}
{(1+e^{\beta (E_p+\tilde{\mu}_q)})^2}\Big].
\end{eqnarray}
So, in the NJL model we have,
\begin{equation}
C_{10}=C_{01}=-2N_fN_c\beta\int\frac{d^3p}{(2\pi)^3}
\Big(\frac{GM}{E_p} \Big)
\Big[\frac{e^{\beta (E_p-{\mu}_q)}}
{(1+e^{\beta (E_p-{\mu}_q)})^2}-
\frac{e^{\beta (E_p+{\mu}_q)}}
{(1+e^{\beta (E_p+{\mu}_q)})^2}\Big].
\label{eq.cross_njl_final}
\end{equation}
Calculation of color trace in PNJL model is already shown in
appendix (\ref{sc.calc_chi}). Using the result of (\ref{eq.tr_final})
and similar one for anti-particle part we arrive at,
\begin{eqnarray}
C_{10}=C_{01}&=&-6 N_f\beta\int\frac{d^3p}{(2\pi)^3}
\Big(\frac{GM}{E_p} \Big)
\Big[
\frac{\bar\Phi e^{-\beta(E_{p}-\mu_q)}
+4\Phi e^{-2\beta(E_{p}-\mu_q)}+3e^{-3\beta(E_{p}-\mu_q)}}
{1+3\bar\Phi e^{-\beta(E_{p}-\mu_q)}
+3\Phi e^{-2\beta(E_{p}-\mu_q)}+ e^{-3\beta(E_{p}-\mu_q)}} 
\nonumber \\
&-& 3 \frac{(\bar\Phi e^{-\beta(E_{p}-\mu_q)}
+2\Phi e^{-2\beta(E_{p}-\mu_q)}+e^{-3\beta(E_{p}-\mu_q)})^2}
{(1+3\bar\Phi e^{-\beta(E_{p}-\mu_q)}
+3\Phi e^{-2\beta(E_{p}-\mu_q)}+ e^{-3\beta(E_{p}-\mu_q)})^2} 
\nonumber \\
&+& \frac{\Phi e^{-\beta(E_{p}+\mu_q)}
+4\bar\Phi e^{-2\beta(E_{p}+\mu_q)}+3e^{-3\beta(E_{p}+\mu_q)}}
{1+3\Phi e^{-\beta(E_{p}+\mu_q)}
+3\bar\Phi e^{-2\beta(E_{p}+\mu_q)}+ e^{-3\beta(E_{p}+\mu_q)}} 
\nonumber \\
&-& 3 \frac{(\Phi e^{-\beta(E_{p}+\mu_q)}
+2\bar\Phi e^{-2\beta(E_{p}+\mu_q)}+e^{-3\beta(E_{p}+\mu_q)})^2}
{(1+3\Phi e^{-\beta(E_{p}+\mu_q)}
+3\bar\Phi e^{-2\beta(E_{p}+\mu_q)}+ e^{-3\beta(E_{p}+\mu_q)})^2}
\Big].
\label{eq.cross_pnjl_final}
\end{eqnarray}
%XXXXXXXXXXXXXXXXXXXXXXXXXXXXXXXXXXXXXXXXXXXXXXXXXXXXXXXXXXXXXXXXXXXXXX%

%----------------------------------------------------------------------%
\subsection{Calculation of R.H.S.\ of Eq.(\ref{eq.delp_delsig2})}
\label{sc.calc_sig2}
%----------------------------------------------------------------------%
\begin{eqnarray}
C_{11}&=&iG^2{\rm Tr}[S_1(P)\cdot S_1(P)] \nonumber \\
&=&-iG^2\int{\frac{d^4P}{(2\pi)^4}
{\rm Tr}_{D,f,c}[S_1(P)\cdot S_1(P)]} \nonumber \\
&=& G^2\frac{4N_f}{\beta}\sum_n\int\frac{d^3p}{(2\pi)^3}\mathrm{Tr}_c
\frac{(i\omega_n+\tilde{\mu}_q)^2
-p^2+M^2}{[(i\omega_n+\tilde{\mu}_q)^2-{E_p}^2]^2}.
\end{eqnarray}
Expressing the integrand in partial fractions we arrive at,
\begin{eqnarray}
C_{11}&=& G^2\frac{N_f}{\beta}\sum_n\int\frac{d^3p}
{(2\pi)^3}\mathrm{Tr}_c
\Big[\Big(\frac{2M^2}{{E_p}^2}\Big)\Big( 
\frac{1}{(i\omega_n+\tilde{\mu}_q-E_p)^2}+\frac{1}
{(i\omega_n+\tilde{\mu}_q+E_p)^2}\Big)
\nonumber \\ &+&\Big(\frac{2p^2}{{E_p}^3}\Big)\Big(
\frac{1}{i\omega_n+\tilde{\mu}_q-E_p}-\frac{1}
{i\omega_n+\tilde{\mu}_q+E_p}\Big)\Big].
\end{eqnarray}
So finally;
\begin{eqnarray}
C_{11} &=& -N_fG^2 ~ \mathrm{Tr}_c\Big[\int\frac{d^3p}{(2\pi)^3}
\Big\{
\frac{2M^2}{{E_p}^2}\beta\Big[
\frac{e^{\beta (E_p-\tilde{\mu}_q)}}
{(1+e^{\beta (E_p-\tilde{\mu}_q)})^2}+
\frac{e^{\beta (E_p+\tilde{\mu}_q)}}
{(1+e^{\beta (E_p+\tilde{\mu}_q)})^2}\Big] \nonumber \\
&-&\frac{2p^2}{{E_p}^3}
\Big[\frac{1}{1+e^{\beta (E_p-\tilde{\mu}_q)}}+
\frac{1}{1+e^{\beta (E_p+\tilde{\mu}_q)}}\Big]\Big\}+
\int_\Lambda\frac{d^3p}{(2\pi)^3}\frac{2p^2}{{E_p}^3}
\Big].
\end{eqnarray}
For NJL model this reduces to,
\begin{eqnarray}
C_{11} &=& -N_cN_fG^2\Big[\int\frac{d^3p}{(2\pi)^3}
\Big\{\frac{2M^2}{{E_p}^2}\beta\Big[
\frac{e^{\beta (E_p-{\mu}_q)}}
{(1+e^{\beta (E_p-{\mu}_q)})^2}+
\frac{e^{\beta (E_p+{\mu}_q)}}
{(1+e^{\beta (E_p+{\mu}_q)})^2}\Big] \nonumber \\
&-&\frac{2p^2}{{E_p}^3}
\Big[\frac{1}{1+e^{\beta (E_p-{\mu}_q)}}+
\frac{1}{1+e^{\beta (E_p+{\mu}_q)}}\Big]\Big\}+
\int_\Lambda\frac{d^3p}{(2\pi)^3}\frac{2p^2}{{E_p}^3}
\Big].
\label{eq.sig2_njl_final}
\end{eqnarray}
Color trace for PNJL model can be done following \cite{ratti_meson}
and using (\ref{eq.tr_final}) which gives,
\begin{eqnarray}
C_{11} &=& -3 N_fG^2\Big[\int\frac{d^3p}{(2\pi)^3}
\Big\{\frac{2M^2}{{E_p}^2}\beta
\Big[
\frac{\bar\Phi e^{-\beta(E_{p}-\mu_q)}
+4\Phi e^{-2\beta(E_{p}-\mu_q)}+3e^{-3\beta(E_{p}-\mu_q)}}
{1+3\bar\Phi e^{-\beta(E_{p}-\mu_q)}
+3\Phi e^{-2\beta(E_{p}-\mu_q)}+ e^{-3\beta(E_{p}-\mu_q)}} 
\nonumber \\
&-& 3 \frac{(\bar\Phi e^{-\beta(E_{p}-\mu_q)}
+2\Phi e^{-2\beta(E_{p}-\mu_q)}+e^{-3\beta(E_{p}-\mu_q)})^2}
{(1+3\bar\Phi e^{-\beta(E_{p}-\mu_q)}
+3\Phi e^{-2\beta(E_{p}-\mu_q)}+ e^{-3\beta(E_{p}-\mu_q)})^2} 
\nonumber \\
&+& \frac{\Phi e^{-\beta(E_{p}+\mu_q)}
+4\bar\Phi e^{-2\beta(E_{p}+\mu_q)}+3e^{-3\beta(E_{p}+\mu_q)}}
{1+3\Phi e^{-\beta(E_{p}+\mu_q)}
+3\bar\Phi e^{-2\beta(E_{p}+\mu_q)}+ e^{-3\beta(E_{p}+\mu_q)}} 
\nonumber \\
&-& 3 \frac{(\Phi e^{-\beta(E_{p}+\mu_q)}
+2\bar\Phi e^{-2\beta(E_{p}+\mu_q)}+e^{-3\beta(E_{p}+\mu_q)})^2}
{(1+3\Phi e^{-\beta(E_{p}+\mu_q)}
+3\bar\Phi e^{-2\beta(E_{p}+\mu_q)}+ e^{-3\beta(E_{p}+\mu_q)})^2}
\Big] \nonumber \\
&-&\frac{2p^2}{{E_p}^3}
\Big[
\frac{\bar\Phi e^{-\beta(E_{p}-\mu_q)}
+2\Phi e^{-2\beta(E_{p}-\mu_q)}+e^{-3\beta(E_{p}-\mu_q)}}
{1+3\bar\Phi e^{-\beta(E_{p}-\mu_q)}
+3\Phi e^{-2\beta(E_{p}-\mu_q)}+ e^{-3\beta(E_{p}-\mu_q)}} 
\nonumber \\
&+& \frac{\Phi e^{-\beta(E_{p}+\mu_q)}
+2\bar\Phi e^{-2\beta(E_{p}+\mu_q)}+e^{-3\beta(E_{p}+\mu_q)}}
{1+3\Phi e^{-\beta(E_{p}+\mu_q)}
+3\bar\Phi e^{-2\beta(E_{p}+\mu_q)}+ e^{-3\beta(E_{p}+\mu_q)}}
\Big] \Big\}
+ \int_\Lambda\frac{d^3p}{(2\pi)^3}\frac{2p^2}{{E_p}^3}
\Big].
\label{eq.sig2_pnjl_final}
\end{eqnarray}
%XXXXXXXXXXXXXXXXXXXXXXXXXXXXXXXXXXXXXXXXXXXXXXXXXXXXXXXXXXXXXXXXXXXXXX%

%%%%%%%%%%%%%%%%%%%%%%%%%%%%%%%%%%%%%%%%%%%%%%%%%%%%%%%%%%%%%%%%%%%%%%%%
%%%%%% BIBLIOGRAPHY %%%%%%%%%%%%%%%%%%%%%%%%%%%%%%%%%%%%%%%%%%%%%%%%%%%%
%%%%%%%%%%%%%%%%%%%%%%%%%%%%%%%%%%%%%%%%%%%%%%%%%%%%%%%%%%%%%%%%%%%%%%%%

%XXXXXXXXXXXXXXXXXXXXXXXXXXXXXXXXXXXXXXXXXXXXXXXXXXXXXXXXXXXXXXXXXXXXXX%

\end{document}